\documentclass[12pt]{article}
\usepackage{graphicx}
\usepackage{epsfig}
\usepackage{rotating}
\usepackage{dcolumn}
\usepackage{bm}
\usepackage{verbatim} 
\usepackage{amsmath}
\usepackage{amssymb}
\usepackage[utf8]{inputenc} 
\usepackage{float}
\usepackage{cite}
\newcommand{\lsim}{\raisebox{-0.13cm}{~\shortstack{$<$ \\[-0.07cm] $\sim$}}~}

\newcommand{\beq}{\begin{eqnarray}} 
\newcommand{\eeq}{\end{eqnarray}} 

\begin{document}

\hfill LPT-Orsay-15-74
\vspace*{1cm} 

\begin{center}

{\large\bf Vector--like top/bottom quark partners and}

\vspace*{.3cm}

{\large\bf  Higgs physics at the LHC}

\vspace*{1cm}

{\sc Andrei Angelescu, Abdelhak Djouadi} and {\sc Gr\'egory Moreau}

\vspace*{.7cm}

Laboratoire de Physique Th\'eorique, Universit\'e Paris--Sud 11 and  CNRS,  

F--91405 Orsay, France.

\end{center}

\vspace*{1cm}

\begin{abstract}

Vector--like quarks (VLQ) that are partners of the heavy top and bottom
quarks are predicted in many extensions of the Standard Model (SM). We
explore the possibility that these states
could explain not only the longstanding anomaly in the forward--backward
asymmetry in $b$--quark production at LEP, \(A_{\rm FB}^b \), but also the
more recent $\sim 2\sigma$ deviation of the cross section for the
associated Higgs production with top quark pairs at
the LHC, $\sigma(pp\to t\bar t H)$. Introducing three illustrative models
for VLQs with
different representations under the SM gauge group, we  show that the two
anomalies can
be resolved while satisfying all other theoretical and experimental
constraints. In this
case, the three different models predict VLQ states in the 1--2 TeV mass
range that can be soon probed at the LHC. In a second step, we discuss the
sensitivity on the VLQ masses and couplings that could be obtained by
means of a percent level accuracy in the measurement of
ratios of partial Higgs decay widths, in particular $\Gamma(H \! \to\!
\gamma\gamma)/\Gamma(H \!  \to\! ZZ^*)$ and $\Gamma(H \! \to \! b\bar
b)/\Gamma(H \! \to \! WW^*)$. We show that top and bottom VL partners with
masses up to $\sim 5$~TeV and exotic VLQs with masses in the 10~TeV range can
be probed at the high--luminosity LHC.
\end{abstract}

\thispagestyle{empty}

\newpage
\setcounter{page}{1}

\subsection*{1. Introduction}

Many extensions of the Standard Model (SM) of particle physics, including some that address the gauge hierarchy problem, predict the existence of additional color-triplet states with vector-like gauge couplings.  Vector-like quarks (VLQs) arise, for instance, as Kaluza-Klein excitations in warped extra-dimension scenarios~\cite{RS} (in particular the version with SM fields in the bulk generating the fermion mass hierarchy, see for example~\cite{RS_bulk_fermions
}), excited resonances in the framework of composite models~\cite{comp_higgs}, partners of the top quark in the little Higgs context~\cite{little_higgs} and as additional states in the extended group representations of grand unified theories~\cite{guts}. As their masses are expected to be in the vicinity of the TeV scale, these particles are accessible at the Large Hadron Collider (LHC) and their search is therefore of prime importance. For phenomenological analyses concerning VLQs, see \cite{vlq_pheno} (for more specific scenarios involving VLQs, see, for example, \cite{vlq_pheno_specific}).

At the LHC, direct experimental searches have imposed the model independent bound $m_{\rm VLQ} \! \gtrsim \! 800$ GeV~\cite{Aad:2015mba,Aad:2015kqa,Aad:2015tba} on VLQ masses from pair-production through strong interactions, almost independently of the electric charge. There exist also indirect constraints on the masses and couplings of these particles from electro-weak (EW) precision tests as they enter the radiative corrections to EW precision observables
such as the so--called oblique corrections that affect the $W$--boson mass $M_W$ and the effective mixing angle $\sin^2\theta_W$ at high orders~\cite{Peskin:1991sw,Altarelli:1990zd}. In addition, third generation VLQs alter the properties of the heavy top and bottom quarks through fermion mixing and strong constraints can be e.g. obtained from the $Z$--boson decay into bottom quarks, $Z\to b\bar b$, as measured at the LEP $e^+ e^-$ collider at energies close to the $Z$--resonance~\cite{Djouadi:1989uk,Boudjema:1989qga}. In the latter case, VLQs are (together with Kaluza--Klein excitations of electroweak gauge bosons \cite{Djouadi:2006rk}) among the very few possibilities that allow to solve the long-standing puzzle of the forward-backward asymmetry $A_{\rm FB}^{b}$ whose measured value differs by $\sim 2.5 \sigma$ from the SM expectation~\cite{Agashe:2014kda}.

Indirect constraints on VLQs also come from the data collected on the 125 GeV Higgs particle that has been observed at the LHC~\cite{Khachatryan:2014jba,Aad:2015gba,HIGGScomb2015}. First, these new quarks contribute to the loop induced Higgs couplings to pairs of gluons and photons, either through their additional exchange in the triangular loops or when altering the important top quark loop contribution by mixing~\cite{Djouadi:2007fm,Azatov:2012rj,Bonne:2012im,Carmi:2012yp,Barger:2012hv}. The Higgs decay channels in the various final states detected so far by the ATLAS and CMS collaborations, namely the $H \to \gamma\gamma,ZZ, WW$ and eventually $\tau^+\tau^-$ final states with the Higgs state dominantly produced in the gluon fusion mechanism $gg\to H$, set strong limits on VLQ masses and couplings~\cite{Khachatryan:2014jba,Aad:2015gba,HIGGScomb2015}. The sensitivity in these leading Higgs production  channels, supplemented by the one in the Higgs--strahlung process $q\bar q \to VH$ with the $V=W,Z$ boson decaying leptonically and the Higgs state decaying into $H\to b\bar b$ final states, will significantly improve at the upgraded LHC with higher center of mass energies and integrated luminosities.

At a later LHC stage, a very efficient indirect probe of VLQ effects would come from associated Higgs production with top quark pairs, $pp\to t \bar tH$, through a modification of the top quark Yukawa coupling $y_t$, as the cross section is directly proportional to $y_t^2$. In fact, the combination of the data collected so far by the ATLAS and CMS collaborations 
in this channel displays a $\sim \! 2\sigma$ deviation from the SM expectation~\cite{HIGGScomb2015} although the sensitivity is still rather low (the deviation is close to $\sim \! 1\sigma$ in the ATLAS data and is much larger, being at the $\sim \! 
2.1\sigma$ level, in the case of CMS~\cite{Khachatryan:2014jba,Aad:2015gba}). This excess in the production rate would correspond to an enhancement of the top--quark Yukawa coupling $y_t$ by a factor $\sim \! 1.4$.\footnote{For an alternative explanation of the $ t \bar tH $ excess, see Ref.~\cite{Huang:2015fba}.} Although it is rather premature, it is tempting to attribute this excess to the indirect presence of VLQs and this should soon be confirmed or infirmed. 

In this paper, we analyze the sensitivity of present and future LHC Higgs data to the vector--like partners of the heavy top and bottom quarks. We adopt an effective approach and consider 
several VLQ representations under the SM gauge symmetry, so that the obtained scenarios can 
be embedded into various realistic high-energy frameworks. We first explore the possibility that some VLQs modify the Yukawa couplings of the heavy top and/or bottom quarks through fermion mixing and discuss the impact of this mixing on electroweak observables including  those in $Z\to b\bar b$ decays. We also analyze the constraints that can be obtained from 
the LHC data on the observed Higgs particle, in particular those from the measured loop 
induced Higgs couplings to gluons and photons as well as from the rates in the Higgs--strahlung production process followed by the decay $H\to b\bar b$. 

As a main outcome of our study, we provide a natural and simultaneous explanation of the two possible deviations in heavy quark observables from SM expectations: the $pp\to t\bar t H$ cross section at the LHC and the $A_{\rm FB}^b$ asymmetry at LEP. For the production rate $\sigma(pp\to t\bar tH)$, the increase of the top Yukawa coupling that is necessary to explain the $ \sim 2\sigma $ excess has to be compensated by a destructive interference between the top and the VLQ loop contributions to the $gg\to H$ production and $H\to \gamma \gamma$ decay rates\footnote{Independently of the present excess in the $pp\to t\bar tH$ production rate, our study provides a motivation for and highlights the importance of a direct measurement of the top quark Yukawa coupling as the indirect determination from the $gg
\! \to \! H$ and $H\! \to \! \gamma \gamma$ processes might be differently altered by new physics.}. Such an interpretation of the anomaly in $\sigma( pp\to t\bar tH)$ predicts VLQs with masses in the range 1--1.5 TeV, which should thus be directly produced at the next LHC runs. 

Finally, we show that VLQs with masses up to $\sim 10$~TeV can be probed by measuring 
precisely the ratios of the $H\! \to \! \gamma\gamma$ to $H\! \to \! ZZ^*$ and $H\! \to \! b \bar b$ to $H\! \to \! WW^*$ production times decay rates~\cite{Djouadi:2012rh,Djouadi:2015aba}, which are free of the large theoretical ambiguities that affect the absolute rates or the signal strengths \cite{Dittmaier:2011ti,Baglio:2010ae,Fichet:2015xla} and which could be determined with an accuracy at the percent level at the high--luminosity LHC option \cite{ATL-PHYS-PUB-2014-016,CMS:2013xfa,ATLAS:2013hta}.

The paper is structured as follows. In the next section, we describe three models which lead to VLQs that could allow for an enhancement of the top quark Yukawa coupling and for a resolution of the $A_{\rm FB}^b$ puzzle. In Section~3, we summarize the presently available constraints that can be set on VLQS, in particular from high precision electroweak and the LHC Higgs data. We then present in Section~4 our numerical results for each studied model and delineate the allowed parameter space for the masses and couplings of VLQs that accommodates the anomalies in $\sigma(pp\to t\bar t H)$  and $A_{\rm FB}^b$. Finally, in Section~5, we discuss the sensitivity to VLQs that can be achieved at the high--luminosity LHC through precision measurements of Higgs decay ratios. A brief conclusion is given in Section~6.

\subsection*{2. The theoretical set-up}
\label{modelbuild} 

In this section, we discuss the simplest models that include extra vector-like quarks and start by analyzing those which could accommodate the two possible anomalies in the heavy quark sector, namely an increase of the $pp\to t\bar tH$ production cross section and a deviation of the $A_{\rm FB}^b$ asymmetry from the SM expectation. In scenarios that lead to modifications of the top quark Yukawa coupling, defined in the mass basis as $y^{\rm SM}_t \! = \! m_t / v$ (when neglecting the three SM generation mixing with respect to the top--VLQ mixing), $m_t\! = 174  \pm 1$~GeV being the measured top--quark mass \cite{Agashe:2014kda} and $v \! = \! v' \sqrt 2  \! \simeq \! 246$~GeV the Higgs vacuum expectation value. The VLQ responsible for such modifications will be denoted as a top partner $t'$ since it should have the same electric charge in order to mix with the top quark. 

It turns out that the simplest SM extension with a unique $t'$ quark leads to a reduction of the top Yukawa coupling with respect to the SM value. This conclusion holds for a $t'$ embedded in a singlet, a doublet or a triplet under the ${\rm SU(2)_L}$ group, because the mass matrix in the $(t,t')$ field basis has the same texture in each of the three cases and generates identical mixing angles. The embedding of a single $t'$ component into a quadruplet or higher ${\rm SU(2)_L}$ multiplets forbids to have gauge invariant Yukawa couplings for the extra $t'$ and, hence, to induce $t$--$t'$ mixing. Consequently, one should include at least two extra top partners. The embedding of vector--like $t',t''$ quarks in two ${\rm SU(2)_L}$ singlets would lead to a mass matrix in the \((t,t',t'')\) field basis of the type (from now on, we denote by ``$Y$" the interaction basis couplings and by ``$y$" the mass basis couplings) 
\begin{equation}
\mathcal{M}_t = \left( \begin{array}{ccc} Y_{t_1} v' & Y_{t_2} v' & Y_{t_3} v' \\ 0 & m_1 & 0 \\ 0 & 0 & m_2 \end{array}\right),
\end{equation} 
which turns out to have an insufficient number of free parameters to increase $y_t$ without significantly altering the measured $m_t$ value. The same holds for two extra isodoublets, for which the mass matrix is simply the transpose of \(\mathcal{M}_t\). Therefore, in order to increase $y_t$, the minimal top sector (where we only consider the least possible number of \(n\)-plets with  \(n \leq 3\)) should include one \(t'\) embedded in an  \(\mathrm{SU(2)_L}\) doublet and one \(\mathrm{SU(2)_L}\) singlet, \(t''\).  In this paper, we consider only 
these minimal scenarios for the heavy quark sectors with the least possible number of \(n\)-plets with  \(n \leq 3\).

In the case of the  forward--backward asymmetry $A_{\rm FB}^b$, one can use similar arguments to construct a minimal sector. The main goal is to reduce the \(A_{\rm FB}^b\) tension with data through tree-level changes of the \(Z b\bar{b}\) couplings, induced by the mixing of the SM $b$--quark with its VLQ partners (see next section). However, one should keep the ratio $R_b \equiv {\Gamma (Z \to b\bar{b})}/{\Gamma (Z \to \mathrm{hadrons})}$ in agreement at the $\lsim 1 \sigma$ level with its SM value when the tree-level \(Z b\bar{b}\) coupling constants \(g_{b_L}\) and \(g_{b_R}\) are modified. This problem has been studied previously  and a possible solution is to increase \(g_{b_R}\) by \( \sim \! 30\%\) and to decrease the absolute value of \(g_{b_L}\) by \( \sim \! 1\%\) with respect to their SM values \cite{Djouadi:2006rk,Bouchart:2008vp}.

The requirement of such a large increase in the right--handed component of the $Zb\bar b$ coupling \(g_{b_R}\) gives an idea on the minimal bottom--quark sector that is required. In the interaction basis, the coupling matrix of the $Z$--boson to the $b$--quark and its VL partners has the diagonal form
\begin{eqnarray}
G_{L/R}^b = \mathrm{diag}(I^{(1)}_{3L/R} + \frac13 \sin^2 \theta_W, \, I^{(2)}_{3L/R} + \frac{1}{3} \sin^2 \theta_W, \, I^{(3)}_{3L/R} + \frac13 \sin^2 \theta_W, \ldots ),
\end{eqnarray}
where \(I^{(1)}_{3R} = 0\) and \(I^{(1)}_{3L} = -\frac12 \) are the SM \(b_L\) and \(b_R\) isospin projections and \(I^{(2,3)}_{3L/R}\) stand for the first and second VL left-handed/right-handed $ b' $'s isospin projections. Rotating to the mass basis by a unitary transformation \(U_R^b\), one finds that, for the characteristic case of two $b'$ states, the \(Z b_R\bar{b}_R\) coupling becomes
\begin{eqnarray}
G_{R,11}^b \equiv \tilde{g}_{b_R} = \sum_{i=1}^{3} I^{(i)}_{3R} \left(U_{R,1i}^b\right)^2 + \frac{1}{3} \sin^2 \theta_W \equiv I^{(1)}_{3R, \rm eff} + \frac{1}{3} \sin^2 \theta_W,
\end{eqnarray}
where \(I^{(1)}_{3R, \rm eff}\) is the ``effective isospin" of the SM bottom quark after mixing with its VLQ partners. Thus, after $b$--$b'$ mixing, the change in \(g_{b_R}\) is equal to \(I^{(1)}_{3R,\rm eff}\), since \(I^{(1)}_{3R} = 0\). As unitarity implies \(\sum_{i=1}^{3} \left(U_{R,1i}^b\right)^2 = 1 \), one concludes that the effective isospin of the SM \(b\) quark is actually a weighted mean of the isospins of all the bottom--like quarks present in the model. Since the measured values of \(A_{\rm FB}^b\) and \(R_b\) point towards \(I^{(1)}_{3R, \rm eff} > 0\),  the minimal model should contain one bottom--like VLQ with positive isospin and none with negative isospin, which from the start excludes a \((t',b')\) doublet.  (Less minimal models could contain additional $b'$ quarks with negative isospins but non-significant mixings with the SM $ b_R $ field, i.e. $ U_{R,1i}^b \ll 1$.) 

Therefore, in the sense of the minimality mentioned above, experimental constraints in the bottom sector
favor a \(b'\) VLQ embedded with a $ -\frac43 $ electric charge VLQ, $ q_{4/3} $, in a \(-\frac56 \) hypercharge isodoublet,
\begin{equation}
B_{L,R} = \left(\!\!\! \begin{array}{c} b' \\ q_{4/3} \end{array} \!\!\! \right)^{ \! Y=-5/6}_{ \! L,R},
\label{Eq:bottom_doublet}
\end{equation}
with the addition of a singlet \(b''\), which guarantees that there are enough parameters to produce a significant deviation of the couplings to the \(Z\) boson \cite{Batell:2012ca}. The electric charge of the multiplet components is fixed by the relation $ Q = Y + I_3 $ coming from the assumption that the symmetry breaking occurs as in the SM. The hypercharge is fixed by the gauge symmetry itself, which imposes the same $ Y $ value for the components of a given multiplet. In addition, as the bottom sector measurements disfavor a \( (t',b') \) doublet and the minimal top sector imposes a \(t'\) embedded in a doublet, one concludes that the \(t'\) should pair up with an exotic electric charge $+\frac53$ VLQ, $ q_{5/3} $, in an \(\rm SU(2)_L\) doublet,
\begin{equation}
T_{L,R} = \left(\!\!\! \begin{array}{c} q_{5/3} \\ t' \end{array} \!\!\! \right)^{ \! Y=7/6}_{ \! L,R}.
\label{Eq:top_doublet}
\end{equation}

Along these lines, one can construct a minimal VLQ model, that we denote here as model {\bf A},  which addresses simultaneously  the excess of the \( pp\to t \bar tH\) cross section  at the LHC and the anomaly in the \(A_{\rm FB}^b\) asymmetry as measured at LEP. Besides the SM fields, model {\bf A} will have the following content:
\begin{equation}
{\rm Model}~\mathbf{A}: T_{L,R}, \; B_{L,R}, \;  t''_{L,R} \; \mathrm{and} \; b''_{L,R},
\end{equation}
where $ B_{L,R} $ and $ T_{L,R} $ are the two isodoublets defined in eqs.~\eqref{Eq:bottom_doublet} and \eqref{Eq:top_doublet} respectively, whereas $ b''_{L,R} $ and $ t''_{L,R} $ are two isosinglets. Denoting the SM left-handed  $(t,b)$ doublet as \(Q_L\), the most general Lagrangian containing all possible terms invariant under the SM ${\rm SU(3)_C \! \times \! SU(2)_L \! \times \! U(1)_Y}$ gauge symmetry reads
\begin{align}
\mathcal{L}   &=         Y_{t_1} \, \overline{Q}_L \tilde{H} \, t_R \, + \, Y_{t_2} \, \overline{Q}_L \tilde{H} \, t''_R 
				\, + \, Y_{t_3} \, \overline{T}_L H \, t_R \, + \, Y_{t_4} \, \overline{T}_L H \, t''_R 
				\, + \, Y_{t_5} \, \overline{T}_R H \, t''_L
\nonumber  \\
  				\, &+ \, Y_{b_1} \, \overline{Q}_L H \, b_R \, + \, Y_{b_2} \, \overline{Q}_L H \, b''_R 
  				\, + \, Y_{b_3} \, \overline{B}_L \tilde{H} \, b_R \, + \, Y_{b_4} \, \overline{B}_L \tilde{H} \, b''_R 
  				\, + \, Y_{b_5} \, \overline{B}_R \tilde{H} \, b''_L 
\nonumber \\
  				\, &+ \, m_1 \, \overline{T}_L T_R \, + \, m_2 \, \overline{t}''_L t''_R 
  				\, + \, m_3 \, \overline{B}_L B_R \, + \, m_4 \, \overline{b}''_L b''_R \, + \, \mathrm{H.c.}, \label{Eq:LagA}
\end{align}
where $ H = \left(\!\!\! \begin{array}{c} H^+ \\ H^0 \end{array} \!\!\! \right) $ represents the SM Higgs doublet, \(\tilde{H}=i \sigma_2 H^*\) its charge conjugate, $L/R$ the left and right fermion chiralities, the $Y$'s dimensionless Yukawa coupling constants and $m$'s the masses of the various VLQs. 

Without loss of generality, the coefficients of the $\overline{t}''_L t_R$ and $ \overline{t}_L t'_R $ terms can be rotated away~\cite{Choudhury:2001hs}. The Yukawa couplings for the first two generations of fermions are omitted in the Lagrangian of eq.~(\ref{Eq:LagA}) as their mixings with the top-partners $t',t''$ are expected to be much smaller than the $t$--$t'$ and $t$--$t''$ mixings as a consequence of the larger mass differences. Since the CKM angles 
\cite{Agashe:2014kda} are typically small, the first two up--quark flavors naturally decouple from the top quark. A similar discussion holds for the down--type quark sector and the $b'$, $b''$ components\footnote{The $t'$ or $b'$ states could contribute to the severely constrained Flavor Changing Neutral Current (FCNC) reactions which rely precisely on the whole SM set of Yukawa couplings for quarks. This issue, which leads to a large number of degrees of freedom in the parameter space, is beyond our scope.}. 

The top Yukawa couplings and mass terms generated after symmetry breaking by the 
Lagrangian of eq.~\eqref{Eq:LagA} can be synthesized respectively in the $h \bar \psi_L^t 
{\cal C}_t \psi_R^t$ and $\bar \psi_L^t {\cal M}_t \psi_R^t$ terms (the ``t" superscript stands for ``top", while the $T$ superscript stands for matrix transposition). Within the interaction basis defined by $ \psi^t=(t,t',t'')^T $, the coupling and mass matrices read
\begin{equation}
{\cal C}_t \ = \  \frac{1}{\sqrt{2}}   \   
\left ( \begin{array}{ccc}
Y_{t_1} & 0 & Y_{t_2} \\  Y_{t_3} & 0 & Y_{t_4} \\ 0 & Y_{t_5} & 0
\end{array} \right )
\   , \   \
{\cal M}_t  \  =  \  
\left ( \begin{array}{ccc}
v'Y_{t_1}  & 0 & v' Y_{t_2} \\  v' Y_{t_3} & m_1 & v' Y_{t_4} \\ 0 & v' 
Y_{t5} & m_2
\end{array} \right ) \ .
\label{ModelMassMat}
\end{equation} 
In the mass basis (``$m$" superscript), one has ${\cal C}_t^m\!=\!U_{L}^t {\cal C}_t (U_{R}^t)^\dagger$, where the unitary matrices $U_{L/R}^t$ are obtained by bi--diagonalizing the model dependent 
mass matrix, $U_L^t {\cal M}_t (U_R^t)^\dagger={\rm diag}(m_{t_1},m_{t_2},m_{t_3})$. The argument stays the same for the $b$--quark sector, but with the replacements $t\to b$, 
\(m_1 \to m_3\) and \(m_2 \to m_4\). As for the $\frac53$ and $-\frac43$ charged 
exotic partners, their masses are given by \(|m_1|\) and \(|m_4|\), respectively. The mass eigenstates obtained in the mass basis are ordered by increasing absolute value and thus, for example, the observed top quark (after mixing) will be represented by \(t_1\), while the lightest bottom-like VLQ will be denoted by \(b_2\).

We will show later that indeed, there is a region in the parameter space of this minimal 
model where all the LEP and LHC constraints, as well as the constraints from the oblique
corrections that affect the $W/Z$ propagators, are satisfied. However, for the sake of completeness, we will also consider two other models that respect too the requirement of minimality and pass the constraints mentioned above. The two additional models contain, besides the SM fields, the VLQ multiplets enlisted below: 
\begin{equation}
{\rm Model}~\mathbf{B}: T_{L,R}, \; B_{L,R}, \;  X_{L,R} = \left( \!\! \begin{array}{c} t'' \\ b'' \\ q'_{4/3} \end{array} \!\! \right)^{\!\!\! Y=-1/3}_{\!\!\! L,R} \! \mathrm{and} \; t'''_{L,R}.
\end{equation}
This is simply a copy of the minimal model {\bf A} with the replacement \( b'' \rightarrow X \), with the top--like singlet from model \textbf{A} being renamed into $ t''' $. The triplet is chosen such that the isospin of $ b'' $ is equal to $0$, which, together with $b'$ having a positive isospin, solves the $ A_{\rm FB}^b $ discrepancy. Also, with the choice of this triplet, this model has the same number of parameters as model \textbf{A}, namely $14$.
\begin{equation}
{\rm Model~}\mathbf{C}: T_{L,R}, \; B_{L,R}, \;  Z_{L,R} = \left( \!\! \begin{array}{c} q_{8/3} \\ q'_{5/3} \\ t'' \end{array} \!\! \right)^{\!\!\! Y=5/3}_{\!\!\! L,R} \!\!, \; b''_{L,R} \;  \mathrm{and} \; t'''_{L,R}.
\end{equation}
Just as in the previous model, the top--like singlet gets the most primes, becoming \(t'''\). In both models, the $ B $ and $ T $ VLQ doublets are the ones defined earlier in eqs.~\eqref{Eq:bottom_doublet} and \eqref{Eq:top_doublet}. 

We close this general discussion by presenting the Lagrangians and the mass matrices of the additional models {\bf B} and {\bf C}. We denote the interaction basis vectors as \(\psi^q = \left(q,q',q'',\ldots\right)^T\), where \(q\) stands for the quark type, namely \(b\), \(t\), \(q_{4/3}\) and \(q_{5/3}\), while $(\ldots)^T$ stands for the matrix transpose operation. 
The Yukawa coupling matrices will not be written, since they are obtained in a straightforward
 manner by differentiating the corresponding mass matrices with respect to the Higgs vev, $v$ (recall that
$v = v' \sqrt 2$).\smallskip

Model \textbf{B}: the corresponding Lagrangian is given by
\begin{align} 
\mathcal{L}   &=         Y_{t_1} \, \overline{Q}_L \tilde{H} \, t_R \, + \, Y_{t_2} \, \overline{Q}_L H \, X_R 
				\, + \, Y_{t_3} \, \overline{Q}_L \tilde{H} \, t'''_R \, + \, Y_{t_4} \, \overline{T}_L H \, t_R 
				\, + \, Y_{t_5} \, \overline{T}_L H \, t'''_R
\nonumber  \\
  				\, &+ \, Y_{t_6} \, \overline{T}_R H \, t'''_L + \, Y_{b_1} \, \overline{Q}_L H \, b_R \, 
  				+ \, Y_{b_2} \, \overline{B}_L \tilde{H} \, b_R \, + \, Y_{b_3} \, \overline{B}_L \tilde{H} \, X_R 
  				\, + \, Y_{b_4} \, \overline{B}_R \tilde{H} \, X_L 
\nonumber \\
  				\, &+ \, m_1 \, \overline{T}_L T_R \, + \, m_2 \, \overline{X}_L X_R 
  				\, + \, m_3 \, \overline{t}'''_L t'''_R \, + \, m_4 \, \overline{B}_L B_R \, + \, \mathrm{H.c.}, \label{Eq:LagB}
\end{align}
For the top, bottom and $-\frac43$ electric charge quarks, the mass matrices are given by 
\begin{gather}
\mathcal{M}_t    =   
\left (\! \begin{array}{cccc}
v'Y_{t_1}  & 0 & v'Y_{t_2} & v'Y_{t_3} \\  v'Y_{t_4} & m_1 & 0 & v'Y_{t_5} 
\\ 0 & 0 & m_2 & 0 \\ 0 & v'Y_{t_6} & 0 & m_3
\end{array} \! \right ), \;
\mathcal{M}_b   =   
\left (\!  \begin{array}{ccc}
v'Y_{b_1}  & 0 & \frac{v' Y_{t_2}}{\sqrt{2}} \\  v'Y_{b_2}  & m_4 & \frac{v'Y_{b_3}}{\sqrt{2}} \\ 0 &\frac{v'Y_{b_4}}{\sqrt{2}} & m_2 
\end{array} \! \right ), \; \notag \\
\mathcal{M}_{4/3}   =   
\left ( \! \begin{array}{cc}
m_4 & v'Y_{b_3} \\ v'Y_{b_4} & m_2
\end{array} \! \right ).
\label{Eq:massB}
\end{gather}
Additionally, the physical mass of \(q_{5/3}\) is given by \( |m1|\). Note that, in the bottom quark mass matrix from above, the \(Y_{t_2}, Y_{b_3}\) and \(Y_{b_4}\) terms are divided by \(\sqrt{2}\). The extra \(1/\sqrt{2}\)'s are just Clebsch-Gordan factors arising from the direct product of the Higgs doublet with a VL doublet into a triplet, i.e. the \(\mathbf{3}\)-representation from the group product decomposition \(\mathbf{2} \otimes \mathbf{2} = \mathbf{3} \oplus \mathbf{1}\).\smallskip

Model \textbf{C}: the Lagrangian is given by
\begin{align} 
\mathcal{L}   &=         Y_{t_1} \, \overline{Q}_L \tilde{H} \, t_R \, + \, Y_{t_2} \, \overline{Q}_L \tilde{H} \, t'''_R 
				\, + \, Y_{t_3} \, \overline{T}_L H \, t_R \, + \, Y_{t_4} \, \overline{T}_L \tilde{H} \, Z_R 
				\, + \, Y_{t_5} \, \overline{T}_L H \, t'''_R
\nonumber  \\
  				\, &+ \, Y_{t_6} \, \overline{Z}_L H \, T_R + \, Y_{t_7} \, \overline{T}_R H \, t'''_L 
  				\, + \, Y_{b_1} \, \overline{Q}_L H \, b_R \, + \, Y_{b_2} \, \overline{Q}_L H \, b''_R 
  				\, + \, Y_{b_3} \, \overline{B}_L \tilde{H} \, b_R 
\nonumber \\
  				\, & + \, Y_{b_4} \, \overline{B}_L \tilde{H} \, b''_R \, + \, Y_{b_5} \, \overline{B}_R \tilde{H} \, b''_L 
  				\, + \, m_1 \, \overline{T}_L T_R \, + \, m_2 \, \overline{Z}_L Z_R 
  				\, + \, m_3 \, \overline{t}'''_L t'''_R 
\nonumber \\  				
  				\, &+ \, m_4 \, \overline{B}_L B_R \, + \, m_5 \, \overline{b}''_L b''_R \, + \, \mathrm{H.c.}, \label{Eq:LagC}
\end{align}
and the mass matrices for the $t, b$ and $\frac{5}{3}$ electric charge quarks are given by
\begin{gather}
\mathcal{M}_t    =   
\left ( \begin{array}{cccc}
v'Y_{t_1} & 0 & 0 & v'Y_{t_2} \\ v'Y_{t_3} & m_1 & v'Y_{t_4} & v'Y_{t_5} \\ 0 & v'Y_{t_6} & m_2 & 0 \\ 0 & v'Y_{t_7} & 0 & m_3
\end{array} \right), \;
\mathcal{M}_b   =   
\left ( \begin{array}{ccc}
v'Y_{b_1} & 0 & v'Y_{b_2} \\  v'Y_{b_3} & m_4 & v'Y_{b_4} \\ 0 & v'Y_{b_5} & m_5 
\end{array} \right), \notag \\
\mathcal{M}_{5/3}   =   
\left ( \begin{array}{cc}
m_1 & \frac{v' Y_{t_4}}{\sqrt{2}} \\ \frac{v' Y_{t_6}}{\sqrt{2}} & m_2
\end{array} \right ).
\label{Eq:massC}
\end{gather}
The novelty of this model is the appearance of an electric charge $\frac83$ exotic quark, \(q_{8/3}\), whose mass is given by \(|m_2|\). Also, the mass of \(q_{4/3}\) is given by \(|m_4|\).

In a later stage of our analysis (when we will exploit the high precision measurement 
of Higgs decay ratios to be performed at the high--luminosity LHC), it would be useful to  simplify to a certain extent the previously considered models in order to keep the discussion as transparent as possible but still at a rather general level. We will thus make the following three simplifying assumptions. 

First, since we would like to study the new physics effects only and not the mixing effects between the SM and the physics beyond it, we will assume the VLQs to decouple from the top and bottom quarks, thus leaving the latter's couplings to the Higgs boson SM-like. This is a good approximation in general since the VLQs that we are investigating have masses well above the electroweak scale and, thus, are supposed to mix weakly with the SM states. At this stage,  we will no longer attempt to explain the LHC hint for an increased top Yukawa coupling nor the anomaly in the \(A_{\rm FB}^b\) asymmetry. 
We will thus allow the new physics that we are considering to communicate with the SM only via the Higgs boson, an assumption which guarantees that the models that we are investigating comply with the currently available phenomenological constraints\footnote{Note that there exist also model--building justifications for such a decoupling of the SM and new physics effects,  such as symmetries canceling the Yukawa coupling terms between SM fields and VLQs.}. 

Second, to focus as much as possible on the effect of a single VLQ and not consider the cumulative contribution of several ones (for instance in the contributions to the \(H \to \gamma\gamma\) or $gg \to H$ loop processes), we will retain for each model only two vector--like multiplets and decouple completely the others. The reason to retain two multiplets and not only one is that at least two fields are needed to have  interactions with the Higgs boson. This interaction with the Higgs field generates, after electroweak symmetry breaking, a mixing term between the two vector--like fields. However, to still concentrate on the effect of a single VLQ, we consider the mass parameter of one of the two multiplets to be larger than the other (this guarantees a small effect of the heavier VLQ  in the loop induced $H\gamma\gamma$ and $Hgg$ vertices for instance). Nevertheless, at the same time, this mass splitting significantly reduces the Yukawa coupling of the lighter VLQ as a result of mixing factors. 

Finally, also for simplicity reasons, we will assume that the two possible Higgs--VLQ--VLQ couplings in the interaction basis are equal, which means that, in the same basis, the VLQ mass matrices are symmetric. The latter have the simple texture
\begin{equation}
\mathcal{M}_{\rm VLQ} = \left( \begin{array}{cc} m & m_Y \\ m_Y & M  \end{array}\right).
\label{Eq:VL_mass_matrix}
\end{equation}
In this expression, $m\;(M$) is the lighter (heavier) VLQ mass parameter, while $m_Y$ is, up to a possible Clebsch--Gordan factor, equal to $ v' Y $ (as  the highest multiplet we consider is a triplet, $1/\sqrt{2}$ is the only possibility for a Clebsch-Gordan coefficient). In each model, $M$ will be fixed to a high value, while $m$ and $Y$ will be treated as variable parameters. 

The choices of the multiplets for the three models introduced before are as follows.

\begin{itemize}
\vspace*{-1mm}

\item For model \textbf{A}, $m$ is the mass parameter of the $ (q_{5/3}, t') $ doublet, while $M$ is the mass parameter of the $t''$ singlet. Here, we will study the effect of the two top partners.\vspace*{-1mm}

\item For model \textbf{B}, $m$ is the mass parameter of the $ (b', q_{4/3}) $ doublet, while $M$ is the mass parameter of the $(t'',b'', q'_{4/3})$ triplet. Here, the main actors will be the exotic $q_{4/3}$ VLQs and, to a much lesser extent, the bottom partners.\vspace*{-1mm} 

\item For model \textbf{C}, $m$ is the VL mass parameter of the $ (q_{5/3}, t') $ doublet, while $M$ is the VL mass parameter of the $(q_{8/3},q'_{5/3}, t'')$ triplet. In this case, we will focus on the contribution of the exotic $q_{5/3}$'s.\vspace*{-1mm}

\end{itemize}

In each of these cases, the eigenmass of the lighter VLQ  will be denoted by $m_{\rm VLQ}$, and its coupling (in the same mass basis) to the Higgs boson by $y_{\rm VLQ}$. These two quantities are deduced from the diagonalization of the matrix in eq.~\eqref{Eq:VL_mass_matrix}. Due to the fact that $M \gg m$, one has $m_{\rm VLQ} \sim m $ and $ y_{\rm VLQ} \sim -v Y^2 / 
(M-m)$.

For the case of the $Hb\bar b$ vertex (which can be probed directly in the measurement of the $H\to b\bar b$ partial width), the discussion will be even simpler. Here, we will consider a non-vanishing mixing between the $b$ quark and its VL partners. In turn, we will consider only bottom-like VLQs since only such states affect the $Hb\bar b$ coupling through $b$--$b'$ mixing. For simplicity, we shall consider the illustrative case of only one bottom-like VL partner $b'$. The choice of the $ \rm SU(2)_L $ representation of the $b'$ extra quark will be qualitatively irrelevant due to similar mass matrix textures (the only quantitative difference could come from various Clebsch-Gordan factors, depending on the $ \rm SU(2)_L $ embedding of $b'$). Thus, a single picture could be representative of all three considered models. For simplicity, we shall take the $b'$ as a singlet under $ \rm SU(2)_L $, which, together with the SM $b$ quark, will lead to a mass matrix given by
\begin{equation}
\mathcal{M}_{b} = \left( \begin{array}{cc}  m_{Y_1} & m_{Y_2} \\ 0 & M  \end{array}\right),
\label{Eq:bot_mass_matrix}
\end{equation}
with $m_{Y_{1,2}} \equiv v' Y_{1,2}$. We shall denote by $y_{b'}$ the Higgs-VLQ coupling in the mass basis and by $m_{b'}$ the bottom-like VLQ eigenmass, both being obtained from the bi-diagonalization of the mass matrix in eq.~\eqref{Eq:bot_mass_matrix}.


\subsection*{3. Present constraints on the VLQ properties}
\label{NumRes} 

\subsubsection*{3.1 Bounds from the LHC Higgs data}
\label{HiggsSector}
 
The first set of constraint that we consider is due to Higgs production and detection
at the LHC; for  a review of the relevant processes see e.g. Ref.~\cite{Djouadi:2005gi}. 
The data collected by the ATLAS and CMS collaborations at $7$+$8$~TeV c.m. energies  in the main search channels, namely the $H\to \gamma\gamma, ZZ, WW, \tau\tau$ detection modes with the Higgs boson produced in the gluon (ggF) and in the vector boson (VBF) fusion channels plus the $H \to b\bar b$ decay mode with the Higgs produced in the $q\bar q \to VH$ mode (VH) 
with $V=W,Z$, seem to be in good agreement with the SM expectations~\cite{Khachatryan:2014jba,Aad:2015gba,HIGGScomb2015}. One can thus use the signal strengths $ \mu_{XX} $ in these Higgs detection channels, defined as the measured cross section times the decay branching ratio relative to the SM prediction,
\begin{equation}
\mu_{XX}= \frac{\sigma( pp\to H)}{\sigma( pp\to H)_{\rm SM}} \times \frac{{\rm BR}(H\to XX)}{{\rm BR}(H\to XX)_{\rm SM}},
\label{Eq:mu_definition}
\end{equation} 
to constrain possible effects of extra vector--like top and bottom partners which would impact several of them. 

The cross section for the gluon fusion mechanism ggF is by far the dominant
Higgs production process at the LHC as it provides $\sim 85\%$ of the total Higgs sample
before kinematical cuts are applied. In the SM, the process is mediated by triangular
top and (to a lesser extent) bottom quark loops. VLQs that are top and bottom partners would affect the ggF production rate either through mixing, i.e. by modifying the $t,b$ loop contributions, or their exchange in the loop (the various quark contributions to the loop-induced $Hgg$ coupling are summarized in the Appendix). The virtual impact of VLQs in the $Hgg$ vertex can be probed essentially through the signal strength in the $H\to ZZ^* \to 4\ell^\pm$ channel that is among  the most precisely measured ones (we refrain here from adding the information from the $H \to WW^* \to 2\ell 2\nu$ search channel that is affected by larger 
theoretical and experimental uncertainties). Averaging the most recent ATLAS and CMS measurements \cite{Khachatryan:2014jba,Aad:2015gba,HIGGScomb2015}, one obtains\footnote{We will not discuss here the subtleties in the treatment of the theoretical uncertainties that are expected to be at the level of 15--20\% in this channel, referring the reader to Ref.~\cite{Dittmaier:2011ti,Fichet:2015xla,Djouadi:2013qya} for a recent discussion
(note that the QCD corrections to the VLQ contributions to the ggF and $H\to \gamma\gamma$ loop processes should be approximately the same as for the top quark contribution; see 
Ref.~\cite{Spira:1995rr} for instance). We also note that a very recent combination of the ATLAS and CMS Higgs results at the first LHC run \cite{HIGGScomb2015} gives slightly different values for the signal strengths in some channels; the difference is nevertheless so small 
that our analysis is unaffected.}~\cite{Khachatryan:2014jba,Aad:2015gba}
\begin{eqnarray}
\mu_{ZZ}^{\rm (comb)} = 1.17^{+0.23}_{-0.22} \label{muZZ} \, .
\end{eqnarray}

The loop induced $H\to \gamma\gamma$ decay mode bears many similarities with the ggF process. It is mediated by top and bottom quark triangular loops but has also contributions from the $W$ boson which, in fact, is dominating and interferes destructively with that of the heavy quarks. Again, additional contributions come from VLQs, in particular through their exchange in the $H\gamma\gamma$ vertex (the impact of VLQs in this channel is also summarized in the Appendix). Given their smaller electric charge, VLQ bottom--quark partners barely contribute to the vertex but exotic VLQs with higher electric charge, e.g. \(+\frac53\) or \(-\frac43\), 
could more significantly affect the loop~\cite{Bonne:2012im}. Present ATLAS and CMS data~\cite{Khachatryan:2014jba,Aad:2015gba}, when combined,  give the even stronger constraint 
\begin{eqnarray}
\mu_{\gamma\gamma}^{\mathrm{(comb)}} = 1.14 \pm 0.18 \, . \label{mupp}
\end{eqnarray}

Additional bottom--like VLQ partners would alter the $Hb\bar{b}$ coupling in addition to the $Z b \bar{b}$ vertex. Consequently, one should also enforce the constraint from the Higgs--strahlung process with the Higgs boson observed in the $H\to b\bar b$ signature. Combining the ATLAS and CMS results~\cite{Khachatryan:2015bnx,Aad:2015gba}, one obtains  for this 
channel\footnote{Here, the theoretical uncertainty is small and the error is largely dominated by the experimental one.}
\begin{eqnarray}
\mu_{bb}^{\rm (comb)} = 0.69 \pm 0.29 \, . \label{mubb}
\end{eqnarray}
Note that here, the production cross section in the VH process is not altered 
at tree level by the presence of VLQ and only the $H\to b\bar b$ branching ratio is 
affected. In fact, this branching ratio, $\sim 60\%$, is the dominant 
one~\cite{Djouadi:1997yw}. It controls the total decay width and therefore enters in all the other Higgs branching 
ratios and hence all signal strengths. We will thus simultaneously include the various
effects and impose the three constraints from  $\mu_{ZZ}, \mu_{\gamma \gamma}$ and 
$\mu_{bb}$ at the same time, ignoring the other signal strengths that are less stringently
constrained~\cite{Khachatryan:2014jba,Aad:2015gba,HIGGScomb2015}.

Finally, we will also consider the signal in the associated $pp\to t\bar tH$ production 
channel for which the combined ATLAS and CMS measurement~\cite{Khachatryan:2014jba,Aad:2015gba,HIGGScomb2015}
\begin{eqnarray}
\mu_{ttH}^{\rm (comb)} = 2.23^{+0.64}_{-0.61} \label{muttH}
\end{eqnarray}
exhibits a $ \sim 2\sigma$ excess compared to the SM value,
that is very tempting  to attribute to new physics. The experimental value for the $\mu_{ttH}$ signal strength assumes SM Higgs decay rates, a feature that is consistent as the decays modified by VLQs such as the $H \to b \bar b$ and $H\to \gamma\gamma$ modes will be separately tested here to be close to their SM values. 

In our discussion, this deviation will be attributed to an enhancement of the top quark Yukawa coupling as a result of mixing with a VLQ partner.  However, because the non--SM--like $y_t$ coupling would also affect the  top--quark contributions to the $Hgg$ and $H\gamma\gamma$ vertices, one could compensate the $y_t $ enhancement by another (negatively interfering) contribution due to VLQ exchanges in the loops as these effective couplings seem
to be in agreement with the SM prediction.

\subsubsection*{3.2 Constraints from high precision tests}
\label{Indirect}

There are also indirect constraints on VLQs from high precision electroweak data. First, for the third generation quark sector, there  are tree-level corrections induced by the $t$--$t'$ or $b$--$b'$ mixings directly on the $t$ or $b$ vertices but, because of the heaviness of $t'$ states, the value for the CKM matrix element $V_{tb}$~\cite{Agashe:2014kda} including quark mixing is expected to be SM--like. In addition, there are radiative corrections to the gauge boson vacuum polarization functions induced by the exchange of VLQs~\cite{Barbieri:2006bg,Lavoura:1992np}. These can be cast into the so--called ``oblique" parameters $S,T$ and $U$~\cite{Peskin:1991sw} that must lie inside the $1\sigma$ regions induced by a long list of electroweak precision observables~\cite{Agashe:2014kda}.  Three crucial observables, the $W$ boson mass $M_W$, the leptonic partial width $\Gamma ( Z\to\ell\ell)$ and the longitudinal polarization and forward--backward asymmetries for leptons that give $\sin^2\theta_W$ play a prominent role~\cite{Agashe:2014kda}. Given the fact that none of our
considered models exhibit an explicit custodial symmetry, it is a non--trivial question whether they will respect these oblique parameter constraints. Also, trying to impose custodial symmetry, as in Ref.~\cite{Batell:2012ca}, would not be a valid solution, since a strong mixing between the SM quarks and the VLQs is needed to explain the $\mu_{ttH}$ enhancement. Moreover, such a symmetry would require a high number of VLQ multiplets, which goes against the idea of minimality.

We will analyze the \(2 \sigma\) excursions of the correlated $S$ and $T$ values (with the usual  assumption that $U = 0$), obtained for our models, from the experimental values of the two parameters, which are given by 
\begin{eqnarray}
S|_{U=0} = 0.06 \pm 0.09~~{\rm and}~~ T|_{U=0} = 0.10 \pm 0.07,
\end{eqnarray}
with a correlation coefficient of \(0.91\)~\cite{Baak:2014ora}. We find that the theoretical prediction for the \(S\) parameter typically does not deviate too much from its central value, while \(T\) has a very high sensitivity to the addition of VLQs. Disentangling the deviations of the observable $T$ that are due to mixing effects or to the VLQ loop contributions is rather difficult in practice. In particular, the mixing effects between (at least) three states are very cumbersome to handle; they can be treated only numerically and one then needs to resort to a scan approach as will be done in our analysis. 

The other set of constraints comes from $Z\to b\bar b$ decays at LEP and one has for the experimental~\cite{Agashe:2014kda} and theoretical~\cite{Agashe:2014kda,Freitas:2014hra} values of the ratio of partial widths $R_b$ and the  asymmetry $A_{\rm FB}^b$, the following values 
\begin{eqnarray}
R_b^{\rm (exp)} = 0.21629 \pm 0.00066 &~ {\rm vs}~& R_b^{\rm (SM)} = 0.2158 
\pm 0.00015 \\
A_{\rm FB}^{\rm b (exp)} = 0.0992 \pm 0.0016 &~ {\rm vs}~& A_{\rm FB}^{b \,(\rm SM)} = 0.1029 \pm 0.0003 
\end{eqnarray}
As already mentioned in several instances, the models that we consider address the $A_{\rm FB}^b$ anomaly. They can be realized within concrete warped extra-dimensional~\cite{Bouchart:2008vp} or their dual composite Higgs scenarios~\cite{Agashe:2006at,DaRold:2010as}. Indeed, in model {\bf A}, the VLQs could be interpreted as Kaluza-Klein excitations of SM quarks in extra-dimensional scenarios. The presence of Kaluza-Klein excitations of the bottom quark would induce $b$--$b'$ mixing and thus corrections to the $Zb\bar b$ couplings that affect $A_{\rm FB}^b$ and $R_b$. Furthermore, extra $t'$ modes would be simultaneously added to enhance the top quark Yukawa coupling. These $t'$ states would then typically have a negative ${\rm SU(2)_L}$ isospin, as explained in Section~2. Such a $t'$ isospin arises in several embeddings in a ${\rm SU(2)_L} \times {\rm SU(2)_R}$ custodial symmetry gauged in the bulk which allows a protection with respect to all electroweak precision data~\cite{Bouchart:2008vp,Agashe:2006at,DaRold:2010as,Agashe:2003zs}. In other words, the extra-dimensional scenarios that comply with the $S,T$  constraints could naturally predict an enhanced $y_t$ coupling and a smaller value for $A_{\rm FB}^b$, at least from the point of view of the field content and their gauge group embedding.

Note that since we are considering a unique set of VLQ fields and not a
replica per generation, it means that the so-called custodians ($t', b', ...$) for the first
two quark (and three lepton) SM generations would decouple, which can be realistic in such frameworks~\cite{Bouchart:2008vp,DaRold:2010as}. Higgs data imply then large masses for the  Kaluza-Klein excitations of gauge bosons and the Higgs sector would essentially feel only the effects of the VLQs (custodians) from the various effective {\bf A--C} models.

\subsubsection*{3.3 Other constraints}

Apart from the constraints coming from the LEP and LHC, one should also incorporate various constraints concerning the eigenmasses of the physical states and their couplings to the scalar Higgs field. 

First, one should reproduce the observed top and bottom quark masses. However, since we are neglecting the mixing between the 3 flavors and also the running from the LHC energy scale down to the heavy quark pole masses $m_t$ and $m_b$, we will allow for an uncertainty for both eigenmasses. In the case of the top quark, $t_1$, we require its mass to lie between $157$ and $ 191 $~GeV, which represents a $10\%$ excursion from the measured value of $m_{t_1} \sim 174$~GeV. As for the bottom quark, $b_1$, we impose for its mass a value between 3 GeV and 5 GeV.

Second, one should take into account the mass constraints coming from direct searches for VLQs at the LHC. Up to date, the most severe bounds on the VLQ masses come from the ATLAS experiment and are as follows:\vspace*{-3mm}

\begin{itemize}
	\item[--] for a top-like VL partner,  $m_{t_2} > 950$~GeV
for $\mathrm{BR} (t_2\to Ht_1) = 1$ \cite{Aad:2015kqa};\vspace*{-2mm} 

	\item[--] for a bottom-like VLQ,   $m_{b_2} > 813$~GeV~for 
$\mathrm{BR} (b_2 \to W t_1) = 1$ \cite{Aad:2015kqa};\vspace*{-2mm}

	\item[--] for a  $+\frac53$ charged VLQ, $m_{5/3_1}>840$~GeV~for 
$\mathrm{BR} (q_{5/3_1}\to Wt_1) = 1$ \cite{Aad:2015mba};\vspace*{-2mm}

	\item[--] for a $-\frac43$ charged VLQ,  $m_{4/3_1} > 770$~GeV~
for $\mathrm{BR} (q_{4/3_1} \to Wb_1) = 1$ \cite{Aad:2015mba}.\vspace*{-2mm}
\end{itemize}

To be conservative, we have considered for each type of VLQ the decay branching ratio values that give the most stringent lower bound on their eigenmass.

Finally, to make our predictions reliable at leading order in perturbation theory, we impose a perturbativity bound on the Yukawa couplings in the mass basis. Using naive dimensional analysis, we thus enforce the conservative constraint $ \max \left( |y_{ij}| \right) < \sqrt{4\pi} $ for all four types of quarks, namely top, bottom, $q_{4/3}$ and $q_{5/3}$ states.

\subsection*{4. Numerical analysis}

We now present our numerical results on the constraints on VLQ masses and couplings from current data. We first summarize the approximations that we use when enforcing the various constraints from the Higgs signal strengths as defined in eq.~\eqref{Eq:mu_definition} and as measured at the first run of the LHC, eqs.~(\ref{muZZ})--(\ref{mubb}).  As discussed previously, on the production side, the additional VLQ can only alter the ggF production mechanism as it does not affect the $HVV$ couplings that enter in the subdominant VBF and VH processes. Since the ggF process is responsible for most of the Higgs production cross section at the LHC, we assume that in both the SM and in our VLQ models, one simply has $\sigma (pp \to H) \simeq \sigma (gg \to H)$  for all signal strengths with the exception of the $ H \to b\bar{b} $ decay. In the latter case, the production mode is instead the Higgs-strahlung process  which should be SM--like.  Moreover, we consider that only the decay $H \to b\bar{b}$, which has the largest branching ratio, is modified in the presence of the VLQs ($H \to \gamma\gamma$ is also modified, but this decay has a negligible branching ratio). Indeed, the other decay mode that involves third generation quark couplings, namely $H\! \to\! gg $, has a small branching ratio and is  expected to be close to its SM value, as the vertex is tested directly via the production process. Thus, we consider that the modification of the total decay width of the Higgs boson, which enters in all signal strengths, comes only from the altered $H b\bar{b} $ vertex.

\begin{figure}[!t]
\begin{center}
\vspace*{1mm}
\mbox{
\hspace*{-.4cm}
\includegraphics[keepaspectratio=true,width=0.48\textwidth]{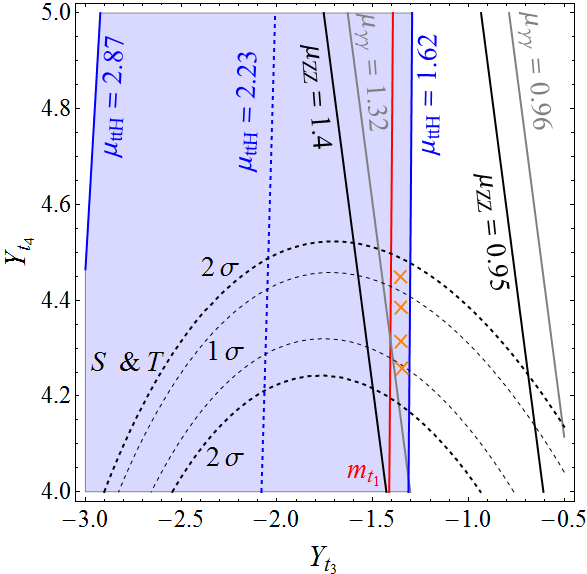}
\includegraphics[keepaspectratio=true,width=0.47\textwidth]{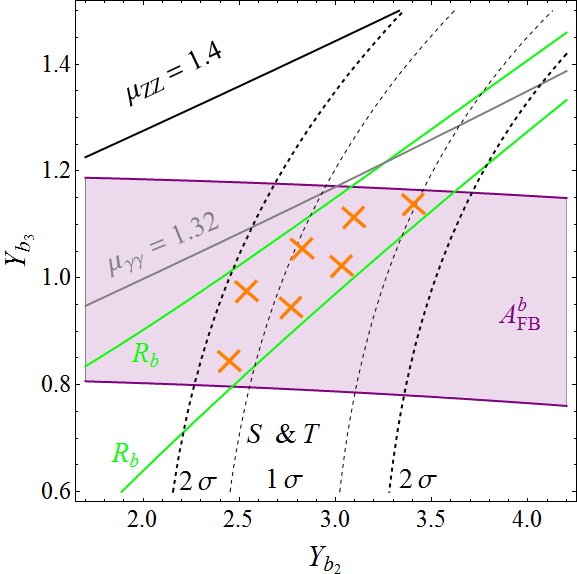}
}
\vspace*{-.4cm}
\caption{{\small In the context of model \textbf{A}, the domains in the $[Y_{t_3},Y_{t_4}]$ plane of the Yukawa couplings in the top sector (left plot) and $[Y_{b_2},Y_{b_3}]$ of the Yukawa couplings in the bottom sector (right plot) in which the various experimental constraints are satisfied at the 1$\sigma$ level. The constraints discussed in the text but not displayed (such as the one from the perturbativity of the Yukawa couplings and the lower limits on the VLQ masses) are satisfied in the entire planes. The regions complying with all constraints are highlighted by the orange crosses.}}
\label{Fig:RegionsA}
\end{center}
\vspace*{-5mm}
\end{figure}

Considering first model \textbf{A}, we present in the left--hand side  of Fig.~\ref{Fig:RegionsA} the constraints that we obtain in the $[Y_{t_3},Y_{t_4}]$ plane. The solid black lines and the grey lines delineate, respectively,  the domains where  the signal strengths  $\mu_{ZZ}$ and  $\mu_{\gamma\gamma}$ respect the LHC measurements given in eqs.~(\ref{muZZ},\ref{mupp}), at the 1$\sigma$ level.  The black dashed lines delineate the areas in which the constraints from the electroweak precision oblique parameter $S$ and $T$ are satisfied at $1\sigma$ and $1\sigma$, with $U=0$, while at the right of the red line, the top quark mass is  reproduced within an uncertainty of $\pm 10\%$ (the lower value does not appear in this frame). The 
regions excluded by the non-perturbativity of the Yukawa couplings or by too low VLQ masses are included but their impact in also not shown in the figure. Finally, the region in which the top quark Yukawa coupling $y_{t_1}$ needs to be enhanced so as to explain the observed excess in the $t \bar t H$ production rate relative to the SM prediction is given by the blue lines: the lines for $\mu_{ttH}= 2.87$ and $\mu_{ttH}= 1.62$, which correspond respectively to the $+1\sigma$ and $-1\sigma$ deviation of the experimental value as given in eq.~(\ref{muttH}), as well as the central value $\mu_{ttH}= 2.23$, are shown. 

Turning to the bottom sector, we display in the right--hand side of Fig.~\ref{Fig:RegionsA}
the regions in the $[Y_{b_2},Y_{b_3}]$  plane 
where the various experimental (and theoretical as we also include the perturbativity of the couplings) constraints are satisfied. Apart from imposing no more than \(1 \sigma\) deviation
compared to the SM for the measured values of the \(A_{\rm FB}^b\) asymmetry (purple lines) and the \(R_b\) ratio of widths (green lines), we allow for the bottom quark mass to take values between \(3\) and \(5\) GeV (to account for the neglected effects of running and flavor mixing), a constraint that is not displayed in the figure as it is satisfied in the entire plane. Also not displayed, the LHC constraint on the $\mu_{bb}$ signal strength is compatible with data at the 1$\sigma$ level in the whole plane (the experimental central value given in eq.~(\ref{mubb}) is $\sim 1\sigma$ smaller than the expectation in the SM). This is not the case of the $ \mu_{ZZ} $ and $ \mu_{\gamma\gamma}$ constraints which, as in the top sector case,  are depicted by the solid black and the solid gray lines respectively. Here, the additional constraints on the mass of the VL bottom--quark partners from direct LHC searches play an important role.  Naturally, the constraints on the top quark mass $m_{t_1}$ and the signal strength $\mu_{ttH}$ do not appear in the plot as they essentially depend on parameters from the top sector (likewise, the constraints from $A_{\rm FB}^b$ and $R_b$ do not appear on the left plot of Fig.~\ref{Fig:RegionsA} as they also do not depend on the top sector parameters).  

In Fig.~\ref{Fig:RegionsA}, the regions with the orange crosses are the ones that are compatible, at the 68\% confidence level (95\% CL for for $S$ and $T$), with all considered constraints. In the bottom sector plot, we fix the $Y_{t_3}$ and $Y_{t_4}$ interaction basis parameters at $Y_{t_3} = -1.45$ and $Y_{t_4} = 4.32$, while in the top sector plot we take $Y_{b_2}=-3.15$ and $Y_{b_3} = -1.08$. The values of the other parameters of the considered model {\bf A} that appear in the Lagrangian of eq.~\eqref{Eq:LagA} are given by \(Y_{t_1} =  -0.98 \), \(Y_{t_2}~\!\!=~\!\!3.05 \), \(Y_{t_5} =  3.81 \), \(Y_{b_1} = -0.02 \),  \(Y_{b_4} = -2.2 \), \(Y_{b_5} = -0.05 \), for the Yukawa couplings and \(m_1  = 1.77\)~TeV, \(m_2  = 1.61\)~TeV, \(m_3  = -0.85 \)~TeV and \(m_4  = - 5.69 \)~TeV for the VLQ masses. 

The outcome of the discussion is that indeed, there is a set of parameters that satisfies 
at the same time the LHC Higgs data and explains in particular the observed excess in the cross section of the $t \bar t H $ process, and the EW precision data accommodating in particular the observed discrepancy of the \(A_{\rm FB}^b\) asymmetry compared to the SM value. As already mentioned, this is a rather non--trivial result.  As the masses of the majority of the VLQs that result from the fit lie in the range between 1 and 2 TeV, this scenario should be soon checked at the LHC by producing directly the additional VLQ states. 

\begin{figure}[!t]
\vspace*{3mm}
\begin{center}
\mbox{
\hspace*{-.3cm}
\includegraphics[keepaspectratio=true,width=0.47\textwidth]{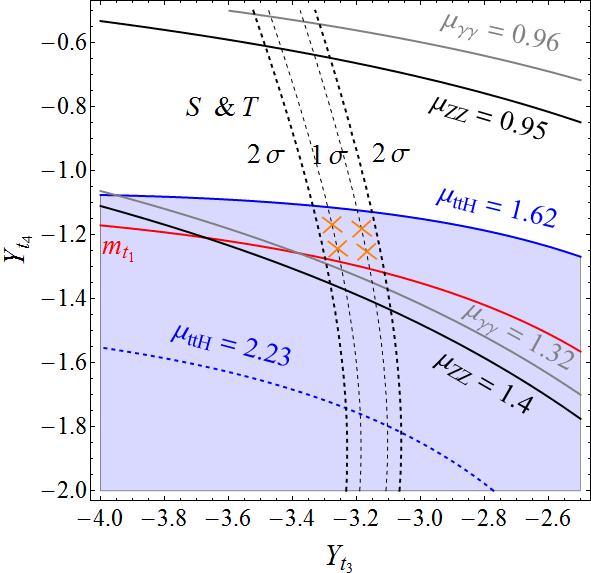}
\includegraphics[keepaspectratio=true,width=0.485\textwidth]{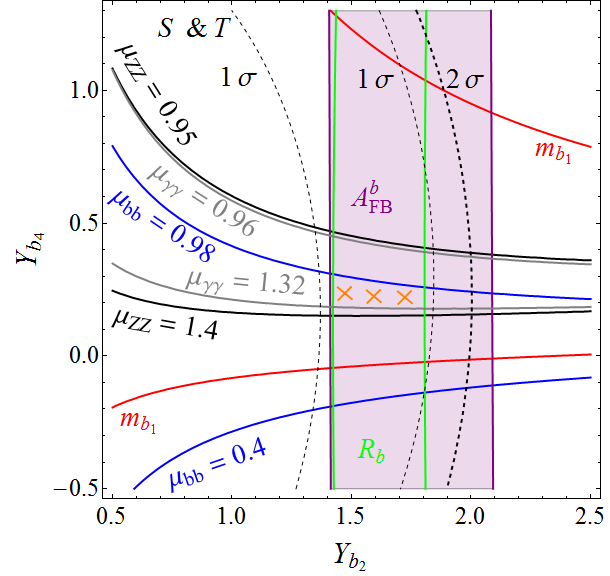}
}
\vspace*{-.2cm}
\caption{{\small The same as in Fig.~\ref{Fig:RegionsA} but in the context of model 
\textbf{B}.
}}
\label{Fig:RegionsB}
\end{center}
\vspace*{-.7cm}
\end{figure}

\begin{figure}[!t]
\vspace*{3mm}
\begin{center}
\mbox{
\hspace*{-.4cm}
\includegraphics[keepaspectratio=true,width=0.48\textwidth]{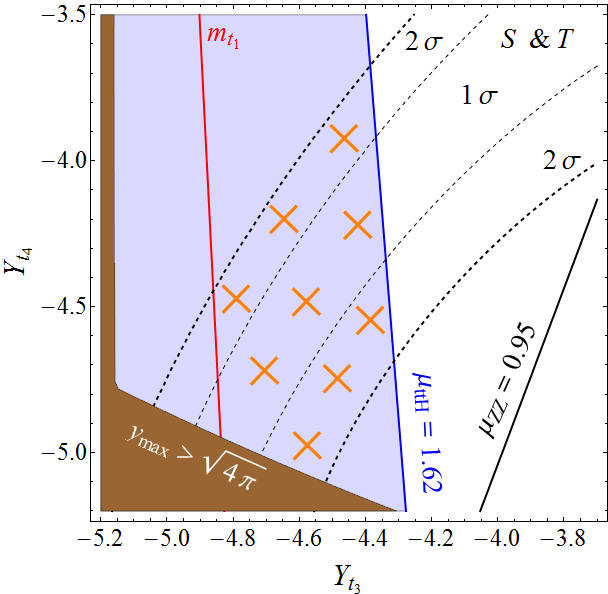}
\includegraphics[keepaspectratio=true,width=0.47\textwidth]{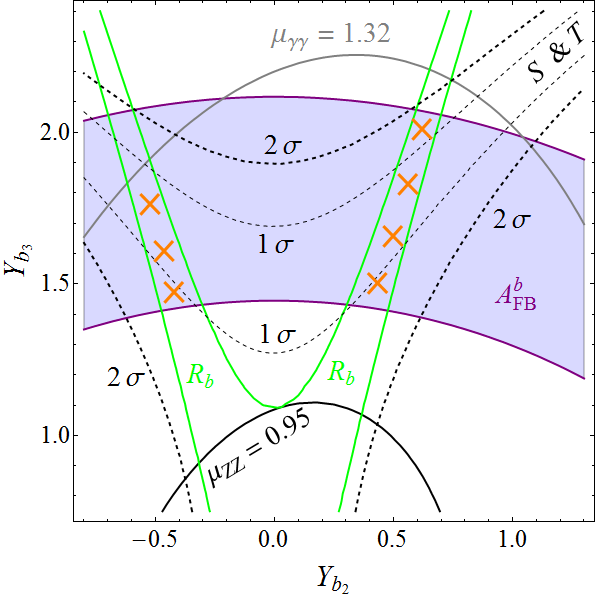}
}
\vspace*{-.3cm}
\caption{{\small The same as in Fig.~\ref{Fig:RegionsA} but in the context of model \textbf{C}. 
}}
\label{Fig:RegionsC}
\end{center}
\vspace*{-0.8cm}
\end{figure}

The same considerations apply for models \textbf{B} and \textbf{C} and we show in 
Fig.~\ref{Fig:RegionsB} and Fig.~\ref{Fig:RegionsC} respectively the impact of the various constraints in the $[Y_{t_3},Y_{t_4}]$ (left plots) and $[Y_{b_2},Y_{b_{3,4}}]$ (right plots) planes. The allowed regions in which the $b$--quark related constraints $m_{b_1}, A_{\rm FB}^b, R_b, \mu_{bb}$, the $t$--quark related constraints $m_{t_1}, \mu_{ttH}$, the general constraints $\mu_{\gamma\gamma}, \mu_{ZZ}$, as well as the EW constraints $S \, \& \, T$ and the lower LHC bounds on the VLQ masses are satisfied at the $1\sigma$ level ($2\sigma$ for $S \, \& \, T$) are also  highlighted by orange crosses. We have also enforced the perturbativity of all Yukawa couplings and the impact of this constraint is now visible in the plots: the dark coloured areas are those where at least one Yukawa coupling in the mass basis is larger than $\sqrt{4\pi}$ in absolute value. Also, we imposed the direct exclusion limit on the VLQ mass $m_{b_2} > 813$ GeV  \cite{Aad:2015kqa}. One constraint is particularly important in the two models \textbf{B} and \textbf{C}, namely the $H \to \gamma\gamma $ signal strength, as both models contain VLQs with high electric charge, $-\frac43$ and $\frac53$, which could lead to important contributions to the $H\gamma\gamma$ vertex. Note that in the left plot of Fig.~\ref{Fig:RegionsC}, only the line for the $-1\sigma$ value of the $t \bar t H $ rate is displayed as we have selected  the areas in which the top Yukawa coupling is sufficiently enhanced to accommodate the observed excess. Moreover, one can see that in the bottom sector plot of model \textbf{C} (right plot of Fig.~\ref{Fig:RegionsC}) there are two disjoint regions where all the phenomenological constraints mentioned above are satisfied. Although they are situated roughly symmetrically with respect to the $ Y_{b_2} = 0 $ line, their shapes are different. This shows that, with all the other parameters fixed, flipping the sign of $ Y_{b_2} $ is of physical importance. Indeed, such a transformation changes the value of $\det \mathcal{M}_b$, which enters directly in the rate expression for the loop-induced ggF mechanism and $H \to \gamma\gamma$ decay (as described in the Appendix).

In the two models, the values of the various parameters appearing (and defined) in eq.~\eqref{Eq:massB} for model \textbf{B} and eq.~\eqref{Eq:massC} for model \textbf{C} and not shown in the planes are as follows. In model {\bf B}, we have: \(Y_{t_1} =  -0.98 \), \(Y_{t_2} = 0.68 \), \(Y_{t_5} =  -3.6 \), \(Y_{t_6} =  4.4 \), \(Y_{b_1} = 0.019 \), \(Y_{b_2} = 1.47 \), \(Y_{b_3} = 0.28 \), \(Y_{b_4} = 0.17 \), \(m_1  = 1.42\)~TeV , \(m_2  = 1.1 \)~TeV, \(m_3  = -2.32 \)~TeV and \(m_4  = 1.5 \)~TeV, with \( Y_{b_2} = 1.61 \) and \( Y_{b_4} = 0.23 \) in the top sector plot (left) plus $ Y_{t_3} = -3.22 $ and $ Y_{t_4} = -1.21 $ in the bottom sector plot (right). In model {\bf C}, we have: \(Y_{t_1} =  -1.01 \), \(Y_{t_2} =  -1.19 \), \(Y_{t_5} = -5.61 \), \(Y_{t_6} = -4.03 \), \(Y_{t_7} = 3.81 \), \(Y_{b_1} = 0.024 \), \(Y_{b_2} = 0.5 \), \(Y_{b_3} = 1.75 \), \(Y_{b_4} = 0.64 \), \(Y_{b_5} = 0.02 \), \(m_1  = -4.8\)~TeV , \(m_2  = -3.12 \)~TeV, \(m_3  = 1.11 \)~TeV, \(m_4  = 1.5 \)~TeV and \(m_5  = 1.1 \)~TeV, with \(Y_{b_2} = 0.52 \) and \(Y_{b_3} = 1.75 \) in the top sector plot (left) plus $ Y_{t_3} = -4.59 $ and $ Y_{t_4} = -4.51 $ in the bottom sector plot (right).

We observe from the three sets of plots Figs.~\ref{Fig:RegionsA}--\ref{Fig:RegionsC} that the allowed regions in the \( [Y_{t_3},Y_{t_4}] \) and the \( [Y_{b_2},Y_{b_3,b_4}] \) planes are not small. Nevertheless, other choices of the remaining parameters do not allow to increase largely those domains. In general, besides \(S\) and \(T\), the most important constraint in the top sector come from enforcing the enhancement of the $t \bar t H $ rate without significantly altering the top quark mass. Simultaneously respecting these two constraints calls for a strong mixing with the extra quarks, which translates into larger Yukawa couplings \( Y_{t_i}\), with \(i=1,2,3,\ldots\). As a consequence, the allowed regions are driven close to the areas ruled out by non-perturbativity, with the highest Higgs--VLQ couplings reaching values typically higher than 3 (for model \textbf{C}, the allowed region only touches the area ruled out by non-perturbativity). Another possibility of enhancing the mixing would be to 
 lower the VLQ mass parameters \(m_i\), but this approach fails, since it leads to VLQ masses that are too low and experimentally excluded. Concerning the bottom sector, the strongest constraints come clearly from the LEP observables \(A_{\rm FB}^b\) and \(R_b\), which are measured at the per mille level, as well as from the \(S\) and \(T\) oblique parameters, measured (indirectly) also at LEP. 
 
Interestingly, the considered models predict the existence of top, bottom ($t_2$ and $b_2$ eigenstates) and even exotic partners around the TeV scale, to which the LHC Run II might be sensitive. While model \textbf{B} predicts 7 VLQs with masses $ \lsim 2 $~TeV, models \textbf{A} and \textbf{C} both predict 4 VLQs with masses $ \lsim 2 $~TeV. Such states will be thus accessible through direct production at the upgraded LHC.

Another feature that can be noticed from  the plots is the fact that in the allowed regions, the top quark mass attains rather large values, usually above \(185\) GeV, while the $t \bar t H $ signal strength has a value around \(1.65\), which is approximately $1\sigma$ below its central value, $2.23$. In fact, the considered VLQ models can more closely reproduce simultaneously the measured top mass \(m_t \simeq 174 \) GeV and a higher value of the $t \bar t H $ signal strength, typically $\mu_{ttH} \simeq 2$ (i.e. only $\sim 0.3 
\sigma$ away from the central value), but at the expense of having \(S\) and \(T\) values outside their \(2 \sigma\) ranges. One can argue that \(S\) and  \(T\), which are measured with a higher accuracy than the Higgs couplings, could be also sensitive to the presence of other sources of new physics, such as extra gauge bosons that appear in many scenarios with VLQs\footnote{This is for instance the case in extra--dimensional models where one would have 
Kaluza--Klein excitations of gauge bosons and top and bottom quark partners. These could contribute to the $S$ and $T$ parameters but not to the Yukawa couplings. Note that the \(A_{\rm FB}^b\) puzzle can be solved by contributions from both extra bosons and/or extra fermions as discussed in Refs.~\cite{Djouadi:2006rk,Bouchart:2008vp}.}, allowing to increase
the range of validity of the Yukawa couplings with the data in Figs.~\ref{Fig:RegionsA}--\ref{Fig:RegionsC}.

Note that the three VLQ models that we  consider improve the discrepancies in  
\(A_{\rm FB}^b\) not only on the $Z$--pole but also off the $Z$--pole. For instance, in model \textbf{A}, for the allowed region of the parameter space in the lower part of Fig.~\ref{Fig:RegionsA}, the \(\chi^2\) function of the fit of all the asymmetry measurements is reduced from \(\chi^2_{\rm SM} \simeq 33 \) down to typically \(\chi^2_{\rm VLQ} \simeq 15 \).

\begin{figure}[!t]
\begin{center}
\hspace*{-4mm}
\includegraphics[keepaspectratio=true,width=1\textwidth]{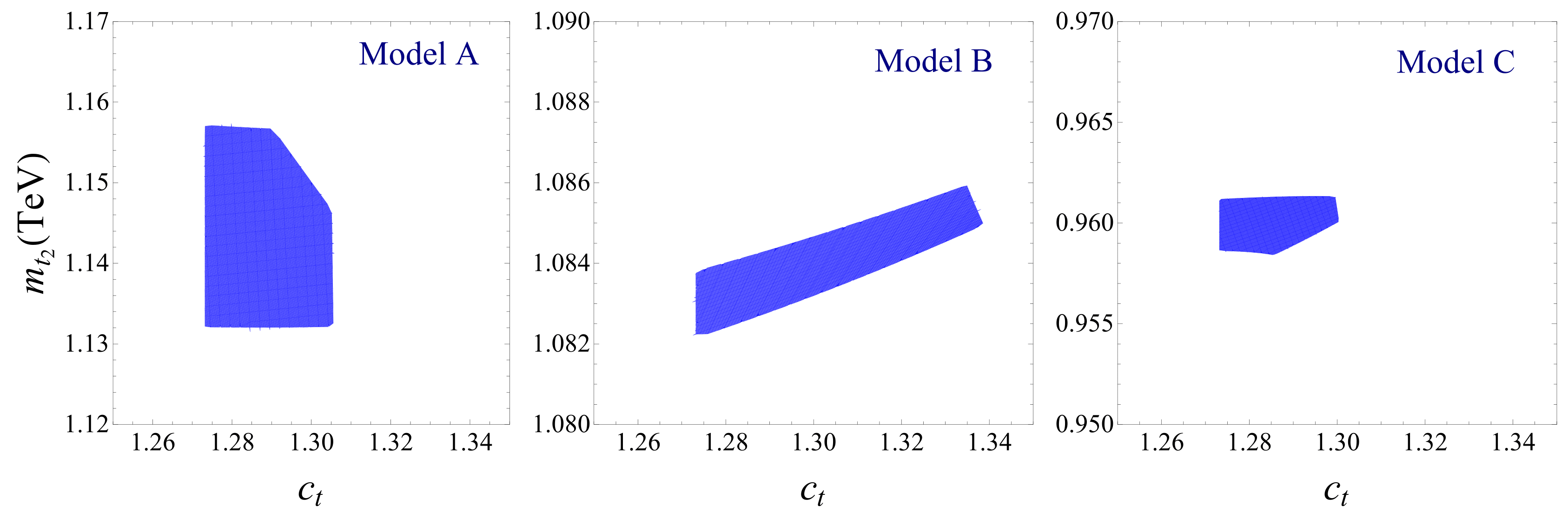}
\includegraphics[keepaspectratio=true,width=1\textwidth]{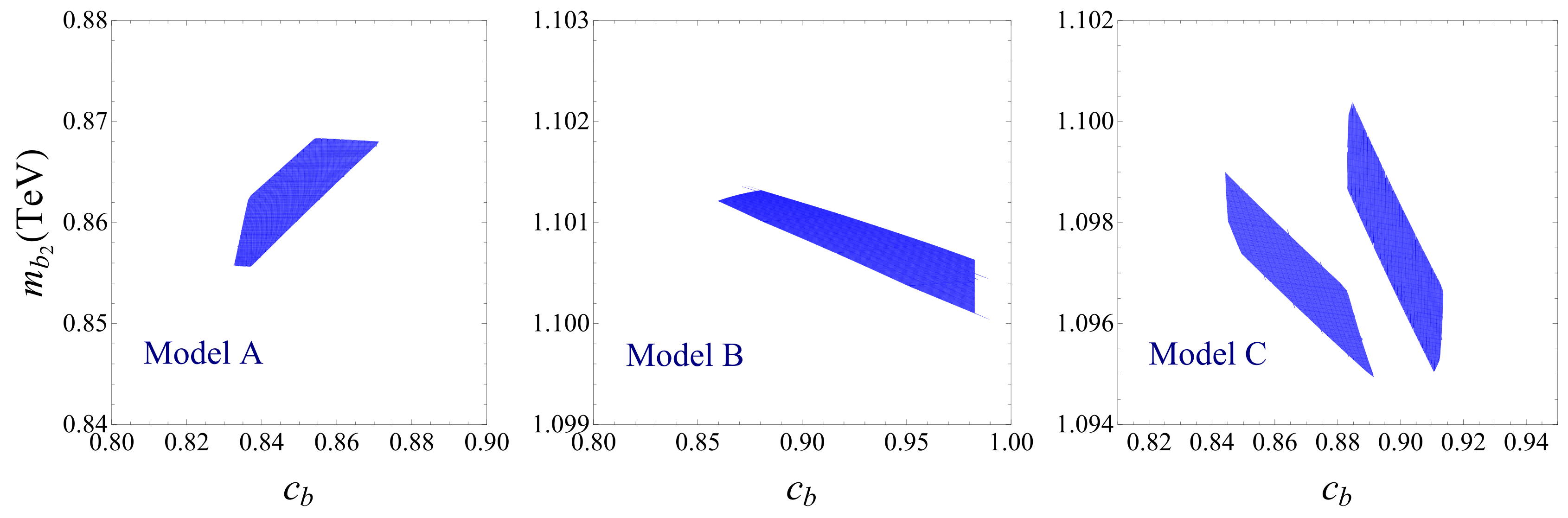}
\vspace*{-8mm}
\caption{
{\small Regions in the $ [ c_{t}, m_{t_2} ] $ plane (upper plots) and the $ [ c_{b}, m_{b_2} ] $ plane (lower plots) where all the phenomenological constraints enlisted in the previous section are satisfied. The varied parameters and their corresponding ranges for the three models are given in the text.}  
}
\label{Fig:summary}
\end{center}
\vspace*{-7mm}
\end{figure}

To summarize the discussion of this section,  we present in Fig.~\ref{Fig:summary} a ``summary plot" containing, for each considered model, the predicted values of $c_{t} \equiv 
|y_{t_1} / y_t^{\rm SM}| $ and $m_{t_2}$ (upper plots) plus $c_{b} \equiv |y_{b_1} / y_b^{\rm SM}| $ and $m_{b_2}$ (lower plots), where $y_{Q_1}$ is the Yukawa coupling (in the mass basis) of the $Q_1$ mass eigenstate, i.e. the observed top and bottom quarks, and 
$y_Q^{\rm SM} $ is the SM prediction (the two values are equivalent in the interaction or mass basis if no fermion mixing is present). Thus, $c_t$ and $c_{b}$ measure the relative departure from the SM of the Yukawa couplings of the top and bottom quarks. As already mentioned throughout the paper, $m_{t_2}$ ($m_{b_2}$) represents the mass of the lightest top--like (bottom--like) VLQ in each of the three obtained models.

In the figure, the varied parameters and their corresponding variation ranges are $ Y_{t_3} \in [-1.6,-1.2] $, $ Y_{t_4} \in [4,4.6] $ and $ Y_{b_2} \in [2,4] $, $ Y_{b_3} \in [0.7,1.2] $ for model~\textbf{A}, $ Y_{t_3} \in [-3.4,-3] $, $ Y_{t_4} \in [-1.4,-1] $ and $ Y_{b_2} \in [1,2] $,  $ Y_{b_4} \in [0,1] $ for model~\textbf{B}, plus $ Y_{t_3} \in [-5,-4.2] $, $ Y_{t_4} \in [-5.5,-3.5] $ and $ Y_{b_2} \in [-0.8,0.8] $, $ Y_{b_3} \in [1.3,2.2] $ for model~\textbf{C}. These intervals cover roughly the allowed regions in Figs.~\ref{Fig:RegionsA}--\ref{Fig:RegionsC} and, for each model, the remaining parameters are fixed at the same values as in these figures. Obviously, the Yukawa couplings with ``$t$" and ``$b$" subscripts correspond respectively to the top and bottom sectors.

The two quantities  $c_t$ and $c_{b}$ are defined as absolute values as the sign of the Yukawa couplings in the mass basis is not physical. Instead, the signs of $y_{t_1}/m_{t_1} $ and $ y_{b_1}/m_{b_1} $ are of physical relevance since such ratios appear directly in the loop-mediated $ gg \to H  $ and $ H \to \gamma\gamma $ amplitudes (see the Appendix
for a discussion). For example, in the $ H \to \gamma\gamma $ process, a negative $y_{t_1}/m_{t_1} $ ratio would mean that the top quark loop amplitude would interfere constructively with the $W$--loop amplitude, leading to an increase of $ \Gamma ( H \to \gamma\gamma) $ with respect to the SM value. In principle, this is possible in general VLQ scenarios but it is not the case in our chosen models. In the regions where all phenomenological constraints are satisfied, we find that in the three models $y_{t_1}/m_{t_1} $ is positive, as in the SM, but slightly higher as a result of the enhancement of the top Yukawa coupling. Depending on the model, the new VLQ mass eigenstates propagating in the loop interfere either constructively or destructively with the top quark exchange. Nevertheless, their contribution to the triangular loop is modest since their masses are rather large and their couplings to the Higgs boson are small, being induced only through quark mixing\footnote{The Higgs--VLQ couplings are given by the diagonal entries of the mass basis Yukawa matrix, $y_{t_{i>1}}$. These entries are zero in the interaction basis so that the mass basis couplings are mixing-induced.}.

As a final remark, there is no complete cancellation between the effects of the enhanced top Yukawa coupling and the contribution of the new VLQ states in the triangular loop. Instead, it turns out that in each of the models that we have considered, the gluon fusion  cross section is increased by 10--15\% compared to the SM value. Meanwhile, relative to its SM value, the diphoton partial width is suppressed by $1-2\%$ in models \textbf{A} and \textbf{B}, whereas in model \textbf{C} it is enhanced by $\sim 15\%$. At present, these slight deviations from the SM are below the experimental accuracy on the signal strengths measured by the ATLAS and CMS collaborations~\cite{Khachatryan:2014jba,Aad:2015gba,HIGGScomb2015}.

\subsection*{5. The sensitivity on VLQs at the upgraded LHC}

We now turn to the discussion of the sensitivity on VLQs that could be achieved at
the upgraded LHC with $\sqrt s=14$ TeV c.m. energy when 3000 fb$^{-1}$ of data will be collected, the so--called high--luminosity option of the LHC (HL-LHC) and start with a
discussion of the observables that can be measured with high precision in this case. 

\subsubsection*{5.1 Precision Higgs observables at high--luminosity}

Compared to $\sqrt s=8$ TeV, the Higgs production cross sections at $\sqrt s=14$ TeV are enhanced by a factor of approximately 2.5 in the case of gluon fusion, 2 in the case of 
Higgs--strahlung and 5 in the case of associated $t \bar t H $ production. The statistical uncertainties on the measurement of the signal strengths values $\mu_{XX}$ for the various processes listed in Section~3.1 and obtained at $\sqrt s=7$+8 TeV with $\sim 25$ fb$^{-1}$ data, will be thus significantly reduced at this LHC upgrade. For instance, in the ggF mode, the statistical error which is presently the largest uncertainty will be reduced by a factor 
$\sqrt {300} \approx 15$ with 3000 fb$^{-1}$ data and would lead to a precision of the order of 1--2\% in the case of the $\mu_{\gamma\gamma}$ and $\mu_{ZZ}$ signal strengths and 3--5\% in the case of $\mu_{bb}$. The smaller systematical uncertainties could also be reduced so that one could hope that the total experimental errors would be reduced to the few percent level in accord with the ATLAS and CMS projection at $\sqrt s=14$ TeV with 3000 fb$^{-1}$ data~\cite{CMS:2013xfa,ATLAS:2013hta}.

The theoretical uncertainties that affect the production cross sections (which are at the level of 10\% in the ggF and 5\% in the VH cases for instance) and the decay branching ratios (which are presently of order 5\% in most channels) would turn then to be the largest source of uncertainty and would limit the interest of these measurements if they are not significantly reduced. Nevertheless, one could construct ratios of observables that are free of these uncertainties. In particular, the ratio of production times decay rates~\cite{Djouadi:2012rh,Djouadi:2015aba}
\begin{align}
D_{\gamma\gamma} &= \frac{\sigma(pp\to H \to \gamma\gamma)}{\sigma(pp\to H \to ZZ^*)} \simeq \frac{\Gamma (H \to \gamma\gamma)}{\Gamma (H \to ZZ^*)}, \\ 
D_{bb} &= \frac{\sigma(q\bar q \to VH \to V b\bar b)}{\sigma(q\bar q \to VH \to V WW^*)} \simeq \frac{\Gamma (H \to b\bar{b})}{\Gamma (H \to WW^*)},
\end{align}
will be free of all these theoretical uncertainties (including also possible ambiguities
in the Higgs total decay width that affect all the branching ratios) provided that the fiducial cross sections for the processes in the numerator and in the denominator are measured within the same kinematical configurations. The two observables will be then limited only by the experimental error and, in particular, the statistical one (at least for $D_{\gamma\gamma}$). At the HL--LHC, one expects that accuracies of the order of   
\beq 
\Delta  D_{\gamma\gamma} \approx 1\% ~~{\rm and}~~ \Delta  D_{bb} \approx 5\%  \label{Dmus}
\eeq
could be achieved. The decay ratios above, which measure only the ratio of Higgs couplings squared $g_{HXX}^2$,  would be then extremely powerful tools to indirectly probe new physics effects and, in particular, those of heavy VLQs of the third generation. 

Another Higgs decay ratio which could also be very useful in general is $D_{\tau \tau}= \Gamma (H \to \tau\tau)/ \Gamma ( H \to WW^*$), with the Higgs state produced in the ggF+1j and VBF
modes. However, we will ignore it in our discussion, since the VLQs that we are analyzing here do not affect the $H\tau\tau$ and $HVV$ couplings and will 
thus have no impact in this context.

Finally, the signal strength in the associated Higgs production with top quark pairs,
$pp\to t\bar tH$, is also important in the context of VLQs. At the HL--LHC, both the ATLAS and CMS collaborations expect a measurement of the cross section $\sigma(pp \to t\bar t H)$ with an experimental accuracy of the order of 15\%~\cite{CMS:2013xfa,ATLAS:2013hta}. This error is largely dominated by the statistical one. In addition to that, the process, which is known at NLO in the QCD and electroweak couplings~\cite{Beenakker:2002nc,Dawson:2003zu,
Frixione:2015zaa,Denner:2015yca}, is affected by a theoretical uncertainty of about 15--20\% from the variation of the renormalisation and factorisation scales and from the parton distribution functions and the value of $\alpha_s$. This leads then to a total uncertainty of about 30\%. Nevertheless, it has been advocated that considering the ratio of cross sections for associated $t \bar t H$ and  $t \bar t Z$ boson production\footnote{Note that in our models, both the $t \bar t H $ and $t \bar t Z $ vertices will be affected via top quark mixing with the VLQs, so that their ratio $C_{tt}$ would not probe solely the $t \bar t H $ vertex.}, $C_{tt}= \sigma(pp\to t\bar t H)/\sigma(pp \to t\bar t Z)$, will also significantly reduce the theoretical uncertainties to the level of $\sim 5\%$ \cite{Plehn:2015cta}. One would then have a total error on the ratio at the level of 15\% when combining the ATLAS and CMS measurements at HL--LHC.

Hence, the ratio $C_{tt} $ is expected to be affected by a much larger error than the $D_{\gamma\gamma}$ and even $D_{bb}$ ratios, thus reducing its capacity to probe tiny VLQ effects. For this reason, although providing a complementary information as it is exclusively sensitive to the $t-t'$ mixing, we will not include this ratio in the rest of our VLQ analysis. 

\subsubsection*{5.2 Probing VLQs using the Higgs decay ratios}

Using the $D_{\gamma\gamma}$ and $D_{bb}$ decay ratios, with the total uncertainties given in eq.~(\ref{Dmus}) and their projected central values equal to their SM values, we now estimate the sensitivities that could be achieved on VLQs at the HL--LHC. We assume that all the measurements of the Higgs couplings at this stage are compatible with the SM expectation and, hence, that  the current anomalies in the $t \bar t H $ production channel and the $b$--quark observables $A_{\rm FB}^b$ and $R_b$ have been resolved  or are ignored. For all the other phenomenological constraints, in particular for the electroweak oblique observables $S$ and $T$, we assume the same central values and errors as presently (we thus ignore for simplicity some potential improvement such as the one that would come from a  better measurement of the $W$ boson mass at the LHC). The ``theoretical" constraints from the top an bottom quark masses and from the perturbativity of the Yukawa couplings, as well as the lower bounds on the masses of the VLQs (which might be improved  by the time of the HL--LHC if no signal is found, but will be superseded by the limits that will be obtained in our analysis) will also be assumed to be the same. 

To simplify the discussion and make the illustration more transparent, we will work in the simplified models presented at the end of Section~2, where only a small subset of the VLQ multiplets with simplified couplings enter Higgs physics. Hence, for the $D_{\gamma\gamma}$ analysis, only two multiplets will be considered, one being light, with a mass parameter $m$, and the other heavy, with a mass parameter $M$. As mentioned at the end of Section~2, both $m$ and $M$ are interaction basis parameters, and the quantities of interest will be the mass of the lighter VLQ mass eigenstate, $m_{\rm VLQ}$, and its mass basis coupling to the Higgs boson, $y_{\rm VLQ}$.  We will assume that, within a few percent approximation, all Higgs couplings (including those to top and bottom quarks) are SM--like and there is almost no mixing in the top and bottom quark sectors between the VLQs and the SM quarks. In this case, only a one percent measurement of the  $D_{\gamma\gamma}$ ratio could signal the new physics effects.   Our goal will be simply to estimate the power of high--precision measurements in the Higgs sector to probe heavy VLQ states with small couplings to the Higgs bosons.

We remind the reader that in the three discussed models, the various multiplets and their impact on the  $D_{\gamma\gamma}$ ratio and hence on the $H\gamma\gamma$ loop are as follows
(as already discussed, VLQ states are also exchanged in the loop induced ggF 
production mechanism but the production rates cancel in the $D_{\gamma\gamma}$ ratio): 
 
\begin{itemize}
\vspace*{-2mm}

\item In model \textbf{A}, $(q_{5/3}, t') $ is the lighter doublet, with mass parameter  
$m$, while the heavier VL field (with the larger mass parameter $M$) is the $t''$ singlet. Both top quark partners will enter in the $H\gamma \gamma$ loop and affect the amplitude.\vspace*{-2mm}  

\item In model \textbf{B}, the $ (b', q_{4/3}) $ doublet is the lighter multiplet while the heavier one is the $(t'',b'', q'_{4/3})$ triplet. Here, the main players will be the exotic $q_{4/3}$ states, while the bottom--quark partners would generate a tiny effect on the triangular Higgs--diphoton loop, of order $(Q_{\rm em}(q_{4/3}) / Q_{\rm em}(b))^2 = 16 $ times smaller than the contribution of the electrically charged $-4/3$ quarks.\vspace*{-2mm}  

\item For model \textbf{C}, the $ (q_{5/3}, t') $ doublet has a mass parameter $m$ and the  $(q_{8/3},q'_{5/3}, t'')$ triplet, a mass $M$. Here, the main contribution will be that of the exotic $q_{5/3}$ states. The contribution of the the top quark partners is approximately $\frac{1}{\sqrt{2}} (Q_{5/3} / Q_{top})^2 \simeq 4.42 $ times smaller than that of the $ q_{5/3} $ states ($ 1/\sqrt{2} $ is a Clebsch-Gordan).\vspace*{-2mm}
\end{itemize}
Some technical aspects concerning our $D_{\gamma\gamma}$ analysis can be found in the Appendix.

As for the $D_{bb}$ decay ratio, we note again that we will consider only one bottom-like VL singlet, $b'$, which will mix with the SM bottom quark and thus alter the $Hb \bar b$ vertex. This simplified scenario is illustrative for all three models we considered. 
In the analysis, we will treat $Y_2$ and $M$, defined in eq.~\eqref{Eq:bot_mass_matrix}, as variable parameters. The remaining parameter, $Y_1$, also appearing in eq.~\eqref{Eq:bot_mass_matrix}, will be expressed in terms of $Y_2$ and $M$ by demanding that $m_b$, the observed $b$ quark mass, is reproduced. Since $ M \gg m_{Y_{1,2}} $ in most of the interesting part of the parameter space, we have, to a very good approximation, $m_b \approx m_{Y_1} (1 - {m_{Y_2}^2}/{2M^2})$, which can easily be inverted in order to re-express $Y_1$ as a function of $Y_2$ and $M$. For this purpose, the value of the bottom quark mass in our numerical analysis will be taken to lie between the $\rm \overline{MS}$ value $m_b (\mathrm{\overline{MS}}) \approx 4.18 $~GeV, and the on-shell value $ m_b(1 \mathrm{S}) \approx 4.65$~GeV~\cite{Agashe:2014kda}, with a mean value $m_b = 4.43$~GeV. Apart from this constraint, we shall enforce the perturbativity condition of the mass basis Yukawa couplings, $y \lesssim \sqrt{4\pi}$, and the LHC bottom-like VLQ exclusion limit, $m_{b'} > 813$~GeV~\cite{Aad:2015kqa}.

\begin{figure}[!t]
\begin{center}
\includegraphics[keepaspectratio=true,width=0.56\textwidth]{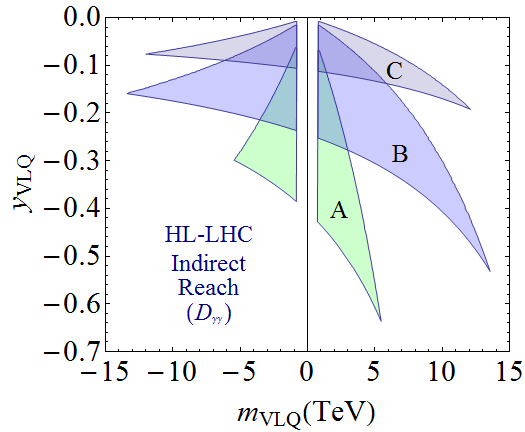}
\vspace*{-.2cm}
\caption{
{\small Regions in the $ [ m_{\rm VLQ},y_{\rm VLQ} ]$ plane for the simplified versions of models \textbf{A}, \textbf{B} and \textbf{C}, to which a precise measurement of $D_{\gamma\gamma}$ at the HL-LHC, with $\Delta D_{\gamma\gamma} = 1\% $,  will be sensitive. The other parameters entering the 
analysis are discussed in the text. }}

\label{Fig:VL_mass_reach}
\end{center}
\vspace*{-7mm}
\end{figure}

We display in Fig.~\ref{Fig:VL_mass_reach}, for the simplified versions of models \textbf{A}-\textbf{C}, regions in the plane $[m_{\rm VLQ},y_{\rm VLQ}]$ to which a precise measurement with $\Delta D_{\gamma\gamma} = 1\% $ will be sensitive. In
this figure, we have assumed that the future central experimental value of $D_{\gamma\gamma}$ would
be equal to its SM prediction. The choices for the heavy VLQ mass parameters are $ M_A = 15 $~TeV, $ M_B = 25 $~TeV and $ M_C = 28 $~TeV. For each model, the ranges of the parameters are $ m \in [-15, 15] $ TeV and $Y \in [0, 5] $. The lower boundary of each region is given by the $ Y = 5 $ curve, which typically marks the transition to the non-perturbativity regime, while the upper boundary is dictated by the $ \Delta D_{\gamma\gamma}  = 1\% $ condition. The region defined by $ \left| m_{\rm VLQ} \right| \lesssim 0.8 $~TeV, delimiting the third boundary, is excluded by direct VLQ searches. 

\begin{figure}[!t]
\begin{center}
\includegraphics[keepaspectratio=true,width=0.56\textwidth]{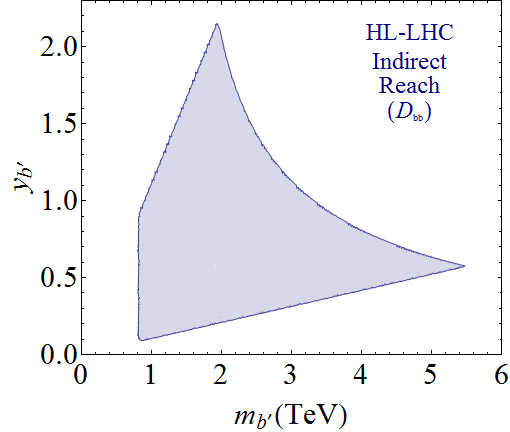}
\vspace*{-.2cm}
\caption{
{\small Regions in the $ [ m_{b'},y_{b'} ]$ plane for the a simplified VLQ model to which a precise measurement of $D_{bb}$ at the HL-LHC, with $\Delta D_{bb} = 5\% $, will be sensitive. The other parameters entering the 
analysis are discussed in the text. }}
\label{Fig:bot_mass_reach}
\end{center}
\vspace*{-7mm}
\end{figure}

Similarly, we present in Fig.~\ref{Fig:bot_mass_reach} regions of the $[m_{b'},y_{b'}]$ plane to which a $5\%$ accuracy measurement of $D_{bb}$ will be sensitive. In
this figure, we have assumed, as in the case of $ D_{\gamma\gamma} $, that the future central experimental value of $D_{bb}$ would
be equal to its SM prediction. Here, the ranges of the parameters are $M \in [0.5, 6]$~TeV and $Y_{b_2} \in [0,6]$. The lower boundary of the obtained region is determined by the $ \Delta D_{bb}= 5\% $ condition, while the upper right one signals the passage to non-perturbativity. The upper left boundary delimits the zone where the observed bottom quark becomes too light, whereas the left boundary shows the lower limit $ m_{b'} \lesssim 0.8$~TeV from direct searches of $b$-like VL partners.

Figs.~\ref{Fig:VL_mass_reach} and \ref{Fig:bot_mass_reach} constitute our main prospective results and one can see that VLQ masses up to several TeV can be probed. With the precise measurement of $D_{\gamma\gamma}$ top partners with masses up to $ 5 $~TeV can be resolved in the loop, while exotic quarks (with $ Q_{\rm em}\! = \! -\! \frac43,  \frac53 $) with masses as high as $\sim 13$~TeV are probed. Meanwhile, a $5\%$ error in the measurement of $D_{bb}$ can be sensitive to the presence of bottom-like VLQs with masses up to $\sim 5$~TeV. It is interesting to observe the complementarity between the two measurements: while with $D_{\gamma\gamma}$ one can efficiently resolve multi-TeV scale top and exotic VL partners,  very heavy bottom VL partners can be probed through $D_{bb}$.

As expected, $ D_{\gamma\gamma} $ is more sensitive to the VLQs with higher electric charge that occur in models \textbf{B} and \textbf{C}. The mass limits above are much higher than the ones obtained from {\it direct} VLQ searches which, even at the HL--LHC, would only reach the 
2 TeV range~\cite{CMS:2013xfa}. It may be surprising that the mass reach for the $ q_{4/3} $'s of model \textbf{B} is higher than the one for the $ q_{5/3} $'s of model \textbf{C}, but the explanation is simple. As it is visible from the figure, model \textbf{C} has a lower sensitivity on $m_{\rm VLQ}$ but for a lower coupling $y_{\rm VLQ}$. The relative smallness of the couplings in model \textbf{C} has two reasons: on the one hand, the Yukawa couplings for model \textbf{C} are suppressed by a Clebsch-Gordan factor of $ 1 / \sqrt{2} $ and, on the other hand, the mass parameter $M$ is larger in model \textbf{C} ($M \sim 28$ TeV) than in model \textbf{B} ($M \approx 25$ TeV), which leads to a smaller mixing between the two VLQs and hence a smaller coupling for the lighter ones to the Higgs boson.

We should also mention that, in the $D_{\gamma\gamma}$ discussion, for models \textbf{B} and \textbf{C}, the oblique parameters $S$ and $T$ are well within $2\sigma$ for all values of $m$ and $Y$ not excluded by non-perturbativity or by direct searches of VLQs. The situation is not as good in model \textbf{A}, where, for $ m_{\rm VLQ} \leq 3 $ TeV, $S$ and $T$ deviate by more than  $3\sigma$. However, since we are interested in knowing the highest possible VL mass that can be resolved in the $ H \to \gamma\gamma$ loop, this is not a serious problem. The case of $D_{bb}$ is similar to the one in model \textbf{A}: for $m_{b'} \gtrsim 3$~TeV, $S$ and $T$ are within $2\sigma$ from their central values.

\subsection*{6. Conclusions}
\label{conclu}

We have analyzed in this paper the sensitivity of present and future LHC Higgs data to heavy  vector--like partners of the top and bottom quarks that appear in many extensions of the SM,
such as warped extra dimension scenarios and composite Higgs models. Working in an effective approach and considering several VLQ representations under the SM gauge group, we have thoroughly investigated three models that address simultaneously the longstanding puzzle of the forward--backward asymmetry \(A_{\rm FB}^b\) at LEP and the recently observed deviation from its SM value of the cross section of the $pp \to t\bar tH$ production process at the LHC. On the other hand, the three models fulfill all other experimental and theoretical constraints, in particular those coming from the electroweak precision measurements and from the LHC data in the Higgs decay and main production channels. 

We have used the principle of minimality as a guide to select representative examples 
of the $t'$ and $b'$ multiplets, which should be related through their contributions to the 
highly constrained electroweak precision data and address the two aforementioned anomalies. 
Among the multiplets that involve $t',b'$ and VLQs with exotic electric charge, one has, for example, $t'$, $b'$ singlets, $(q_{5/3},t')$, $(b', q_{4/3})$ doublets and/or $(t',b',q_{4/3})$, $(q_{8/3}, q_{5/3},t')$  triplets. These states mix with the SM top and bottom quarks and modify their Yukawa and gauge couplings. In addition, they would contribute to the loop induced $gg\to H$ production and $H\to \gamma\gamma$ decay processes. For instance, the mixing with the additional states in the bottom sector allows for a sufficiently large increase of the \(Z b_R b_R\) coupling to explain the \(A_{\rm FB}^b\) anomaly. At the same time, an enhancement of the $Ht\bar t$ Yukawa coupling by a factor up
to $\sim 1.4$ can occur,  which would instead explain the $\sim 2\sigma$ apparent increase of the cross section $\sigma(pp\to t\bar t H)$ at the LHC. The rates for the 
loop induced processes would stay SM--like due to either small VLQ contributions or compensating effects between
fermion mixing and  loop contributions. Interestingly, the considered models predict the existence of VLQ with masses in the range 1--2 TeV  that might be discovered at the current Run II of the LHC with a c.m. energy of 13 to 14 TeV. 

In a second part of the paper, we left aside the anomalies in the asymmetry $A_{\rm FB}^b $ and the cross section $\sigma(pp \to t\bar{t}H) $ and focused instead on the VLQ mass scale  that could be probed in the future by precision measurements in the Higgs sector at the high--luminosity LHC option. In this context, the ratios of the partial widths of the $H\to \gamma \gamma$ vs $H\to ZZ^*$ and $H\to b\bar b$ vs $H\to WW^*$ decay modes,  $D_{\gamma\gamma} $ and $D_{bb}$, would play an important role as they can be determined with an accuracy at the level of, respectively,  $\Delta D_{\gamma\gamma} = 1\%$ and $\Delta D_{bb} = 5\%$. Assuming the worst-case scenario in which the new physics scale would lie far above the electroweak scale and all other measured observables would appear to be SM-like, we have shown that, in some simplified VLQ frameworks, the precise measurement of the two decay ratios would probe VLQs with masses above the multi--TeV range. In particular, VLQs contributing to the the $H\gamma\gamma$ loop vertex or altering at tree--level the $Hb\bar b$ coupling would be visible at the HL-LHC if the mass scales are $ \sim 5 $~TeV for top and bottom partners and up to $\sim 13$~TeV for VLQs with higher electric charge, such as $-\frac43$ or $ \frac53$. These mass values are much higher than those attainable in  direct VLQ searches at the LHC in the present~\cite{Aad:2015mba,Aad:2015kqa,Aad:2015tba} or even in the future~\cite{CMS:2013xfa,ATLAS:2013hta}.\smallskip

\noindent {\bf Acknowledgements:}
A.D. would like to the CERN Theory Unit for the hospitality extended to him. 
This work is supported by the ERC advanced grant {\it Higgs@LHC}. G.M. also thanks support from the ``Institut Universitaire de France'' and the European Union FP7 ITN INVISIBLES (Marie Curie Actions, PITN-GA-2011-289442).

\newpage
\subsection*{Appendix: VLQ contributions to the $Hgg$ and $H\gamma\gamma$ vertices}

Noting the Yukawa couplings for the VL mass eigenstates as $y_{t_i}$ and \(y_{b_i}\), (labeled by $i=1,2,\ldots$), the ratio of the ggF cross section over its SM 
prediction reads as,
\begin{equation}
\mu_{\rm ggF} \equiv \frac{\sigma^{\rm VL}_{\rm ggF}}{\sigma_{\rm ggF}^{\rm SM}} \ = \ 
\frac{\big \vert \sum_i \ \frac{v \,  y_{t_i}}{m_{t_i}} A[\tau(m_{t_i})] +  \sum_i \ \frac{v \,  y_{b_i}}{m_{b_i}} A[\tau(m_{b_i})] \big \vert^2}{\big \vert A[\tau(m_{t})] + A[\tau(m_{b})]\big \vert^2}  \, \tag{A.1}
\label{Eq:ggF1}
\end{equation}
where $A[\tau(m)]$ is the form factor for spin~1/2 particles~\cite{Djouadi:2005gi} 
normalized such that $A[\tau(m)\ll 1]\to \frac43$ and $A[\tau(m)\gg 1]\to 0$ with $\tau(m)=m_H^2/4m^2$. It is useful to use this large mass limit, which is
a reasonable approximation (except for the bottom quark), so that the first sum from eq.~(\ref{Eq:ggF1}) simplifies~\cite{Frank:2013un}, 
\begin{equation}
\sum_i \ \frac{v  y_{t_i}}{m_{t_i}} A[\tau(m_{t_i})] \simeq  \sum_i   \frac{v  y_{t_i}}{m_{t_i}} 
 = v \, {\rm Tr}(\frac{\partial {\cal M}_t}{\partial v} {\cal M}^{-1}_t)  
 = v \frac{\partial}{\partial v} \log \det {\cal M}_t . \tag{A.2}
\label{Eq:ggF2}
\end{equation}
Eq.~\eqref{Eq:ggF2} is useful as, due to the invariance of the trace with respect to basis changing, it can be applied to the matrix $\mathcal{M}_t$ in the starting interaction basis as well, thus avoiding the explicit calculation of the basis transformation. A similar trick can be used for the sum over the bottom quark states. However, since the bottom form factor is almost zero, one has to add and then subtract its contribution:
\begin{equation}
\sum_i \ \frac{v  y_{b_i}}{m_{b_i}} A[\tau(m_{b_i})] \simeq \sum_i \frac{v  y_{b_i}}{m_{b_i}} - \frac{v  y_{b_1}}{m_{b_1}} 
= v \frac{\partial}{\partial v} \log \det {\cal M}_b - \frac{v  y_{b_1}}{m_{b_1}}, \tag{A.3}
\label{Eq:ggF3}
\end{equation}
which enables us to write an approximate form for the ggF ratio from eq.~\eqref{Eq:ggF1}:
\begin{equation}
\mu_{\rm ggF} \simeq v^2 \left| \frac{\partial}{\partial v} \log \det {\cal M}_t + \frac{\partial}{\partial v} \log \det {\cal M}_b - \frac{y_{b_1}}{m_{b_1}} \right|^2. \label{Eq:ggF4} \tag{A.4}
\end{equation}
Additionally, in order to use the above equation as a transparent guide to start the exploration of a multivariate parameter space, one can take the \( v y_{b_1} / m_{b_1} \) term to be \(\mathcal{O}(1)\), just as in the SM. Moreover, if there are other higher charged exotic quarks that couple to the Higgs boson, a term of the type \(\frac{\partial}{\partial v} \log \det {\cal M}_X\), with \(X\) denoting the quark type, should be added to the previous equation. Thus, eq.~\eqref{Eq:ggF4} can be used as a guide to stay in regions where the ggF rate is not too far from its SM value. Nevertheless, in our numerical analyses, we use the exact expression of eq.~(\ref{Eq:ggF1}).

As for the \(R_{\gamma\gamma}\) ratio (which is different from $D_{\gamma\gamma})$,
\begin{equation}
R_{\gamma\gamma} \equiv \frac{\Gamma(H \to \gamma\gamma)}{\Gamma(H \to \gamma\gamma)_{SM}}, \tag{A.5}
\end{equation}
which enters in $\mu_{\gamma\gamma}$, a similar formula can be derived \cite{Djouadi:2005gi}, and the tricks displayed above can be used once again. The difference is that the $W$--boson also runs in the loop, which generates an additional term besides the ones in eq.~\eqref{Eq:ggF4}, and that each sum over quarks gets multiplied by \(N_C = 3\) and by \(\left(Q_{e.m.}(q)\right)^2\), where \(Q_{e.m.}(q)\) is the \(\mathrm{U(1)_{em}}\) charge of the quark in question. Therefore, the bottom quarks' contribution becomes negligible, hence allowing one to reliably estimate \(R_{\gamma\gamma}\) by differentiating with respect to \(v\) the various \(\log\det \mathcal{M}_q\) terms.

We turn now to some  technical aspects concerning the study of $D_{\gamma\gamma}$ in the context of our simplified VLQ models. The values for the interaction basis parameters ($Y$, $m$ and $M$) and for the mass basis quantities ($ m_{\rm VLQ}$ and $y_{\rm VLQ}$) were chosen on the following grounds. First of all, $Y$ was chosen to be always positive because its sign has no impact on the physics. To understand more easily the nature of $Y$'s sign, let us denote the lighter VLQ as $q$ and the heavier one as $Q$. There exists a $ \mathbb{Z}_2 $ field transformation under which, for example, $q$ is odd and $Q$ is even. Performing such a transformation on the two VLQ fields, which leaves the VLQ kinetic and mass terms invariant, would change the sign of the $ Y (\bar{Q} q + \mathrm{H.c.})$ terms of the Lagrangian. By virtue of this transformation, $Y$ can always be set to a positive value, which means that its sign is not physically meaningful. Also, we have chosen $Y \leq 5$ because, for $Y \gtrsim 5$, at least one of the four Yukawa couplings in the mass basis becomes non-perturbative, i.e. greater than $\sim \sqrt{4\pi}$. Moreover, while $m$ was allowed to attain both negative and positive values, $M$ was chosen to be positive for all the models, since only the relative sign of the two parameters influences the physics. 

The explanation reads as follows. Since we assume that there is no mixing between the SM fields and the VLQs, we have for the amplitude of the $H \to \gamma\gamma$ process
\begin{equation}
\mathcal{A}_{\gamma\gamma} = \mathcal{A}_{\rm SM} + \mathcal{A}_{\rm VLQ}, \tag{A.6}
\label{Eq:D_gg}
\end{equation} 
where $\mathcal{A}_{\rm SM}$ and $\mathcal{A}_{\rm VLQ}$ are respectively the SM state-mediated and VLQ-mediated loop amplitudes. As explained in the beginning of this appendix, in the limit where the VLQs are much heavier than the Higgs boson,
\begin{equation}
\mathcal{A}_{\rm VLQ} \propto \frac{\partial}{\partial v} \left[ \log \left( \det\mathcal{M}_{\rm VLQ} \right) \right].
\tag{A.7}
\end{equation}
Thus, $A_{\gamma\gamma}$ depends only on $ \frac{\partial}{\partial v} \left[ \log \left( \det\mathcal{M}_{\rm VLQ} \right) \right] $ and, therefore, the sign of a parameter is physically relevant only if it affects the VLQ mass matrix determinant, which is given by
\begin{equation}
\det\mathcal{M}_{\rm VLQ} = mM - m_Y^2. \tag{A.8}
\label{Eq:det_vlq}
\end{equation}
This expression shows that only the sign of $mM$ enters in the studied observable quantity and confirms that the sign of $Y$ is unphysical (recall that $ m_Y \propto Y $). Finally, the values of $M$ for each model were taken such that the largest resolvable $m_{\rm VLQ}$ is roughly half of $M$, which avoids too much feedback in the $H \to \gamma\gamma$ loop from the heavier VLQ and thus isolates to some extent the contribution to $D_{\gamma\gamma}$ of the lighter VLQ.

As for the mass basis quantities, $ y_{\rm VLQ} $ and $ m_{\rm VLQ}$, the discussion is somewhat simpler. Only their relative sign is of physical importance, since they enter $D_{\gamma\gamma}$ through their ratio, as it appears for example in  eq.~\eqref{Eq:ggF1}  for the comparable structure of the ggF loop amplitude. Therefore, we plotted only negative values for $y_{\rm VLQ}$, while letting $m_{\rm VLQ}$ have any sign. Nevertheless, we did not plot the region where $\left| m_{\rm VLQ} \right| \leq 0.8 $~TeV (the empty band along the $m_{\rm VLQ}=0$ line), since it is excluded by direct searches for VL partners.

Also, it is interesting to note in Fig.~\ref{Fig:summary} the asymmetry of the regions with respect to the $ m_{\rm VLQ} = 0 $ axis. This is due to the fact that, in the case of $ m_{\rm VLQ} > 0 $, the interference of the lighter VLQ with its heavier counterpart is destructive, while for $ m_{ \rm VLQ} < 0 $ the exact opposite happens. Thus, the values of $ y_{\rm VLQ} $ that can be probed are higher in the case of a positive mass for the lighter VLQ. Loosely speaking, the situation is the other way around if the sign of $ M $ is flipped. More precisely, under the change $ M \rightarrow -M $, both $ m_{\rm VLQ} $ and $ y_{\rm VLQ} $ would change sign, which graphically means that the regions in Fig.~\ref{Fig:VL_mass_reach} would undergo a reflection about the origin, defined by $ \{ m_{\rm VLQ}, y_{\rm VLQ} \} = \{ 0,0 \}$. 

\newpage

\begin{small}

\end{small}


\begin{thebibliography}{999} 

\bibitem{RS}
  M.~Gogberashvili,
  Int.\ J.\ Mod.\ Phys.\ D {\bf 11} (2002) 1635
  [hep-ph/9812296];
  L.~Randall and R.~Sundrum,
  Phys.\ Rev.\ Lett.\  {\bf 83} (1999) 3370
  [hep-ph/9905221].
\vspace{-2mm}


\bibitem{RS_bulk_fermions}
  T.~Gherghetta and A.~Pomarol,
  Nucl.\ Phys.\ B {\bf 586} (2000) 141
  [hep-ph/0003129];
  S.~J.~Huber and Q.~Shafi,
  Phys.\ Lett.\ B {\bf 512} (2001) 365
  [hep-ph/0104293];
  G.~Moreau and J.~I.~Silva-Marcos,
  JHEP {\bf 0603} (2006) 090
  [hep-ph/0602155];
  G.~Moreau and J.~I.~Silva-Marcos,
  JHEP {\bf 0601} (2006) 048
  [hep-ph/0507145];
  Y.~Grossman and M.~Neubert,
  Phys.\ Lett.\ B {\bf 474} (2000) 361
  [hep-ph/9912408];
  G.~Moreau,
  Eur.\ Phys.\ J.\ C {\bf 40} (2005) 539
  [hep-ph/0407177].
\vspace{-2mm}


\bibitem{comp_higgs}
  R.~Contino, Y.~Nomura and A.~Pomarol,
  Nucl.\ Phys.\ B {\bf 671} (2003) 148
  [hep-ph/0306259];
  K.~Agashe, R.~Contino and A.~Pomarol,
  Nucl.\ Phys.\ B {\bf 719} (2005) 165
  [hep-ph/0412089].
\vspace{-2mm}


\bibitem{little_higgs}
  N.~Arkani-Hamed, A.~G.~Cohen and H.~Georgi,
  Phys.\ Lett.\ B {\bf 513} (2001) 232
  [hep-ph/0105239];
  N.~Arkani-Hamed, A.~G.~Cohen, E.~Katz and A.~E.~Nelson,
  JHEP {\bf 0207} (2002) 034
  [hep-ph/0206021].
  \vspace{-2mm}


\bibitem{guts}
  J.~Kang, P.~Langacker and B.~D.~Nelson,
  Phys.\ Rev.\ D {\bf 77} (2008) 035003
  [arXiv:0708.2701 [hep-ph]];
  C.~Kilic, K.~Kopp and T.~Okui,
  Phys.\ Rev.\ D {\bf 83} (2011) 015006
  [arXiv:1008.2763 [hep-ph]].
\vspace{-2mm}


\bibitem{vlq_pheno}

  F.~del Aguila, J.~A.~Aguilar-Saavedra and R.~Miquel,
  Phys.\ Rev.\ Lett.\  {\bf 82} (1999) 1628
  [hep-ph/9808400];
  J.~A.~Aguilar-Saavedra,
  Phys.\ Lett.\ B {\bf 625} (2005) 234
   [Phys.\ Lett.\ B {\bf 633} (2006) 792]
  [hep-ph/0506187];
  J.~A.~Aguilar-Saavedra,
  JHEP {\bf 0911} (2009) 030
  [arXiv:0907.3155 [hep-ph]];
  G.~Cacciapaglia, A.~Deandrea, L.~Panizzi, N.~Gaur, D.~Harada and Y.~Okada,
  JHEP {\bf 1203} (2012) 070
  [arXiv:1108.6329 [hep-ph]];
  S.~Gopalakrishna, T.~Mandal, S.~Mitra and R.~Tibrewala,
  Phys.\ Rev.\ D {\bf 84} (2011) 055001
  [arXiv:1107.4306 [hep-ph]];
  G.~Moreau,
  Phys.\ Rev.\ D {\bf 87} (2013) 1,  015027
  [arXiv:1210.3977 [hep-ph]];
  G.~Cacciapaglia, A.~Deandrea, L.~Panizzi, S.~Perries and V.~Sordini,
  JHEP {\bf 1303} (2013) 004
  [arXiv:1211.4034 [hep-ph]];
  M.~Buchkremer, G.~Cacciapaglia, A.~Deandrea and L.~Panizzi,
  Nucl.\ Phys.\ B {\bf 876} (2013) 376
  [arXiv:1305.4172 [hep-ph]];
  J.~A.~Aguilar-Saavedra, R.~Benbrik, S.~Heinemeyer and M.~Pérez-Victoria,
  Phys.\ Rev.\ D {\bf 88} (2013) 9,  094010
  [arXiv:1306.0572 [hep-ph]];
  S.~Gopalakrishna, T.~Mandal, S.~Mitra and G.~Moreau,
  JHEP {\bf 1408} (2014) 079
  [arXiv:1306.2656 [hep-ph]];
  S.~Dawson and E.~Furlan,
  Phys.\ Rev.\ D {\bf 89} (2014) 1,  015012
  [arXiv:1310.7593 [hep-ph]];
  C.~Y.~Chen, S.~Dawson and I.~M.~Lewis,
  Phys.\ Rev.\ D {\bf 90} (2014) 3,  035016
  [arXiv:1406.3349 [hep-ph]];
  G.~Cacciapaglia, A.~Deandrea, N.~Gaur, D.~Harada, Y.~Okada and L.~Panizzi,
  JHEP {\bf 1509} (2015) 012
  [arXiv:1502.00370 [hep-ph]];
  N.~Vignaroli,
  Phys.\ Rev.\ D {\bf 91} (2015) 11,  115009
  [arXiv:1504.01768 [hep-ph]];
  C.~Y.~Chen, S.~Dawson and Y.~Zhang,
  Phys.\ Rev.\ D {\bf 92} (2015) 7,  075026
  [arXiv:1507.07020 [hep-ph]];
  F.~del Aguila, M.~Perez-Victoria and J.~Santiago,
  Phys.\ Lett.\ B {\bf 492} (2000) 98
  [hep-ph/0007160];
  F.~del Aguila, M.~Perez-Victoria and J.~Santiago,
  JHEP {\bf 0009} (2000) 011
  [hep-ph/0007316];
  A.~Atre, M.~Carena, T.~Han and J.~Santiago,
  Phys.\ Rev.\ D {\bf 79} (2009) 054018
  [arXiv:0806.3966 [hep-ph]];
  A.~Atre, G.~Azuelos, M.~Carena, T.~Han, E.~Ozcan, J.~Santiago and G.~Unel,
  JHEP {\bf 1108} (2011) 080
  [arXiv:1102.1987 [hep-ph]];
  R.~Barcelo, A.~Carmona, M.~Chala, M.~Masip and J.~Santiago,
  Nucl.\ Phys.\ B {\bf 857} (2012) 172
  [arXiv:1110.5914 [hep-ph]];
  A.~Atre, M.~Chala and J.~Santiago,
  JHEP {\bf 1305} (2013) 099
  [arXiv:1302.0270 [hep-ph]];
  S.~Fajfer, A.~Greljo, J.~F.~Kamenik and I.~Mustac,
  JHEP {\bf 1307} (2013) 155
  [arXiv:1304.4219 [hep-ph]];
  A.~K.~Alok, S.~Banerjee, D.~Kumar, S.~U.~Sankar and D.~London,
  Phys.\ Rev.\ D {\bf 92} (2015) 013002
  [arXiv:1504.00517 [hep-ph]];
  A.~K.~Alok, S.~Banerjee, D.~Kumar and S.~U.~Sankar,
  arXiv:1402.1023 [hep-ph];
  N.~Bizot and M.~Frigerio,
  arXiv:1508.01645 [hep-ph].
\vspace{-2mm}


\bibitem{vlq_pheno_specific}
  P.~Lodone,
  JHEP {\bf 0812} (2008) 029
  [arXiv:0806.1472 [hep-ph]];
  J.~P.~Araque, N.~F.~Castro and J.~Santiago,
  arXiv:1507.05628 [hep-ph];
  M.~Gillioz, R.~Gröber, A.~Kapuvari and M.~Mühlleitner,
  JHEP {\bf 1403} (2014) 037
  [arXiv:1311.4453 [hep-ph]];
  N.~Maru and N.~Okada,
  Phys.\ Rev.\ D {\bf 87}, no. 9, 095019 (2013)
  [arXiv:1303.5810 [hep-ph]];
  N.~Maru and N.~Okada,
  arXiv:1310.3348 [hep-ph];
  J.~Carson and N.~Okada,
  arXiv:1510.03092 [hep-ph];
C.~Han, A.~Kobakhidze, N.~Liu, L.~Wu and B.~Yang,
Nucl.\ Phys.\ B {\bf 890}, 388 (2014)
[arXiv:1405.1498 [hep-ph]];
N.~Liu, L.~Wu, B.~Yang and M.~Zhang,
arXiv:1508.07116 [hep-ph];
   R.~N.~Mohapatra and Y.~Zhang,
   JHEP {\bf 1406}, 072 (2014)
   [arXiv:1401.6701 [hep-ph]].
\vspace{-7mm}
  

\bibitem{Aad:2015mba}
  G.~Aad {\it et al.} [ATLAS Collaboration],
  Phys.\ Rev.\ D {\bf 91} (2015) 11,  112011
  [arXiv:1503.05425 [hep-ex]].
\vspace{-2mm}

\bibitem{Aad:2015kqa}
  G.~Aad {\it et al.} [ATLAS Collaboration],
  JHEP {\bf 1508} (2015) 105
  [arXiv:1505.04306 [hep-ex]].
\vspace{-7mm}

\bibitem{Aad:2015tba}
  G.~Aad {\it et al.} [ATLAS Collaboration],
  arXiv:1509.04261 [hep-ex].
\vspace{-2mm}


\bibitem{Peskin:1991sw}
  M.~E.~Peskin and T.~Takeuchi,
  Phys.\ Rev.\ D {\bf 46} (1992) 381.
\vspace{-2mm}
\bibitem{Altarelli:1990zd}
  G.~Altarelli and R.~Barbieri,
  Phys.\ Lett.\ B {\bf 253} (1991) 161.
\vspace{-2mm}


\bibitem{Djouadi:1989uk}
  A.~Djouadi, J.~H.~Kuhn and P.~M.~Zerwas,
  Z.\ Phys.\ C {\bf 46} (1990) 411.
\vspace{-2mm}
\bibitem{Boudjema:1989qga}
  F.~Boudjema, A.~Djouadi and C.~Verzegnassi,
  Phys.\ Lett.\ B {\bf 238} (1990) 423.
\vspace{-2mm}

\bibitem{Djouadi:2006rk}
  A.~Djouadi, G.~Moreau and F.~Richard,
  Nucl.\ Phys.\ B {\bf 773} (2007) 43
  [hep-ph/0610173].
\vspace{-2mm}

\bibitem{Agashe:2014kda}
  K.~A.~Olive {\it et al.} [Particle Data Group Collaboration],
  Chin.\ Phys.\ C {\bf 38} (2014) 090001.
\vspace{-2mm}


\bibitem{Khachatryan:2014jba}
  V.~Khachatryan {\it et al.} [CMS Collaboration],
  Eur.\ Phys.\ J.\ C {\bf 75} (2015) 5,  212
  [arXiv:1412.8662 [hep-ex]].
\vspace{-2mm}
\bibitem{Aad:2015gba}
  G.~Aad {\it et al.} [ATLAS Collaboration],
  arXiv:1507.04548 [hep-ex].
\vspace{-2mm}
\bibitem{HIGGScomb2015} ATLAS and CMS Collaborations, ATLAS-CONF-2015-044.\vspace{-2mm}


\bibitem{Huang:2015fba}
  P.~Huang, A.~Ismail, I.~Low and C.~E.~M.~Wagner,
  Phys.\ Rev.\ D {\bf 92} (2015) 7,  075035
  [arXiv:1507.01601 [hep-ph]].
\vspace{-2mm}


\bibitem{Djouadi:2007fm}
  A.~Djouadi and G.~Moreau,
  Phys.\ Lett.\ B {\bf 660} (2008) 67
  [arXiv:0707.3800 [hep-ph]].
  \vspace{-2mm}
\bibitem{Azatov:2012rj}
  A.~Azatov, O.~Bondu, A.~Falkowski, M.~Felcini, S.~Gascon-Shotkin, D.~K.~Ghosh, G.~Moreau and S.~Sekmen,
  Phys.\ Rev.\ D {\bf 85} (2012) 115022
  [arXiv:1204.0455 [hep-ph]].
  \vspace{-7mm}
\bibitem{Bonne:2012im}
  N.~Bonne and G.~Moreau,
  Phys.\ Lett.\ B {\bf 717} (2012) 409
  [arXiv:1206.3360 [hep-ph]].
\vspace{-2mm}
\bibitem{Carmi:2012yp}
  D.~Carmi, A.~Falkowski, E.~Kuflik and T.~Volansky,
  JHEP {\bf 1207} (2012) 136
  [arXiv:1202.3144 [hep-ph]].
  \vspace{-2mm}
\bibitem{Barger:2012hv}
  V.~Barger, M.~Ishida and W.~Y.~Keung,
  Phys.\ Rev.\ Lett.\  {\bf 108} (2012) 261801
  [arXiv:1203.3456 [hep-ph]].
\vspace{-2mm}


\bibitem{Djouadi:2012rh}
  A.~Djouadi,
  Eur.\ Phys.\ J.\ C {\bf 73} (2013) 2498
  [arXiv:1208.3436 [hep-ph]].
\vspace{-2mm}
\bibitem{Djouadi:2015aba}
  A.~Djouadi, J.~Quevillon and R.~Vega-Morales,
  arXiv:1509.03913 [hep-ph].
  \vspace{-2mm}


\bibitem{Dittmaier:2011ti}
  S.~Dittmaier {\it et al.} [LHC Higgs Cross Section Working Group Collaboration],
  arXiv:1101.0593 [hep-ph].
\vspace{-2mm}
\bibitem{Baglio:2010ae}
  J.~Baglio and A.~Djouadi,
  JHEP {\bf 1103} (2011) 055
  [arXiv:1012.0530 [hep-ph]].
\vspace{-2mm}
\bibitem{Fichet:2015xla}
  S.~Fichet and G.~Moreau,
  arXiv:1509.00472 [hep-ph].
\vspace{-2mm}


\bibitem{ATL-PHYS-PUB-2014-016} ATLAS Collaboration, ATL-PHYS-PUB-2014-016.\vspace{-2mm}
\bibitem{CMS:2013xfa} CMS Collaboration, CMS-NOTE-13-002, arXiv:1307.7135 [hep-ex].\vspace{-2mm}
\bibitem{ATLAS:2013hta} ATLAS Collaboration, ATL-PHYS-PUB-2013-007, arXiv:1307.7292 [hep-ex].\vspace{-2mm}


\bibitem{Bouchart:2008vp}
  C.~Bouchart and G.~Moreau,
  Nucl.\ Phys.\ B {\bf 810} (2009) 66
  [arXiv:0807.4461 [hep-ph]].
\vspace{-2mm}

\bibitem{Batell:2012ca}
  B.~Batell, S.~Gori and L.~T.~Wang,
  JHEP {\bf 1301} (2013) 139
  [arXiv:1209.6382 [hep-ph]].
\vspace{-2mm}

\bibitem{Choudhury:2001hs}
  D.~Choudhury, T.~M.~P.~Tait and C.~E.~M.~Wagner,
  Phys.\ Rev.\ D {\bf 65} (2002) 053002
  [hep-ph/0109097].
\vspace{-2mm}

\bibitem{Djouadi:2005gi}
  A.~Djouadi,
  Phys.\ Rept.\  {\bf 457} (2008) 1
  [hep-ph/0503172].
\vspace{-2mm}

\bibitem{Djouadi:2013qya}
  A.~Djouadi and G.~Moreau,
  Eur.\ Phys.\ J.\ C {\bf 73} (2013) 9,  2512
  [arXiv:1303.6591 [hep-ph]].
  \vspace{-7mm}

\bibitem{Spira:1995rr}
  M.~Spira, A.~Djouadi, D.~Graudenz and P.~M.~Zerwas,
  Nucl.\ Phys.\ B {\bf 453} (1995) 17
  [hep-ph/9504378].
\vspace{-2mm}

\bibitem{Khachatryan:2015bnx}
  V.~Khachatryan {\it et al.} [CMS Collaboration],
  Phys.\ Rev.\ D {\bf 92} (2015) 3,  032008
  [arXiv:1506.01010 [hep-ex]].
  \vspace{-2mm}

\bibitem{Djouadi:1997yw}
  A.~Djouadi, J.~Kalinowski and M.~Spira,
  Comput.\ Phys.\ Commun.\  {\bf 108} (1998) 56
  [hep-ph/9704448].
\vspace{-2mm}


\bibitem{Barbieri:2006bg}
  R.~Barbieri, L.~J.~Hall, Y.~Nomura and V.~S.~Rychkov,
  Phys.\ Rev.\ D {\bf 75} (2007) 035007
  [hep-ph/0607332].
  \vspace{-2mm}
\bibitem{Lavoura:1992np}
  L.~Lavoura and J.~P.~Silva,
  Phys.\ Rev.\ D {\bf 47} (1993) 2046.
\vspace{-2mm}

\bibitem{Baak:2014ora}
  M.~Baak {\it et al.} [Gfitter Group Collaboration],
  Eur.\ Phys.\ J.\ C {\bf 74} (2014) 3046
  [arXiv:1407.3792 [hep-ph]].
  \vspace{-2mm}

\bibitem{Freitas:2014hra}
  A.~Freitas,
  JHEP {\bf 1404} (2014) 070
  [arXiv:1401.2447 [hep-ph]].
\vspace{-2mm}


\bibitem{Agashe:2006at}
  K.~Agashe, R.~Contino, L.~Da Rold and A.~Pomarol,
  Phys.\ Lett.\ B {\bf 641} (2006) 62
  [hep-ph/0605341].
\vspace{-2mm}
\bibitem{DaRold:2010as}
  L.~Da Rold,
  JHEP {\bf 1102} (2011) 034
  [arXiv:1009.2392 [hep-ph]].
\vspace{-2mm}
\bibitem{Agashe:2003zs}
  K.~Agashe, A.~Delgado, M.~J.~May and R.~Sundrum,
  JHEP {\bf 0308} (2003) 050
  [hep-ph/0308036].
\vspace{-2mm}


\bibitem{Beenakker:2002nc}
  W.~Beenakker, S.~Dittmaier, M.~Kramer, B.~Plumper, M.~Spira and P.~M.~Zerwas,
  Nucl.\ Phys.\ B {\bf 653} (2003) 151
  [hep-ph/0211352].
\vspace{-2mm}
\bibitem{Dawson:2003zu}
  S.~Dawson, C.~Jackson, L.~H.~Orr, L.~Reina and D.~Wackeroth,
  Phys.\ Rev.\ D {\bf 68} (2003) 034022
  [hep-ph/0305087].
\vspace{-2mm}
\bibitem{Frixione:2015zaa}
  S.~Frixione, V.~Hirschi, D.~Pagani, H.-S.~Shao and M.~Zaro,
  JHEP {\bf 1506} (2015) 184
  [arXiv:1504.03446 [hep-ph]].
\vspace{-2mm}
\bibitem{Denner:2015yca}
  A.~Denner and R.~Feger,
  arXiv:1506.07448 [hep-ph].
\vspace{-2mm}

\bibitem{Plehn:2015cta}
  M.~L.~Mangano, T.~Plehn, P.~Reimitz, T.~Schell and H.~S.~Shao,
  arXiv:1507.08169 [hep-ph].
\vspace{-7mm}

\bibitem{Frank:2013un}
  M.~Frank, N.~Pourtolami and M.~Toharia,
  Phys.\ Rev.\ D {\bf 87} (2013) 9,  096003
  [arXiv:1301.7692 [hep-ph]].
  
\end{thebibliography}
\end{document}